\documentclass{article}

\PassOptionsToPackage{round}{natbib}
\usepackage[final]{pccl_paper}

\usepackage{amsmath}

\usepackage[utf8]{inputenc} 
\usepackage[T1]{fontenc}    
\usepackage{hyperref}       
\usepackage{cleveref}       
\usepackage{url}            
\usepackage{booktabs}       
\usepackage{amsfonts}       
\usepackage{nicefrac}       
\usepackage{microtype}      
\usepackage{xcolor}         
\usepackage{graphicx} 
\usepackage{tikz}           
\usepackage{listings}       

\usepackage{algorithm}
\usepackage{algpseudocode}
\usepackage{algorithmicx}
\usepackage{caption}
\usepackage{booktabs}
\usepackage{tabularx}

\usepackage{pgfplots}
\pgfplotsset{compat=1.17}

\usetikzlibrary{arrows.meta, shapes, positioning}

\newcommand{\pccl}{\textsc{PCCL}}

\title{Prime Collective Communications Library - Technical Report}

\author{
  Michael Keiblinger \\
  Prime Intellect \\
\And
  Mario Sieg \\
  Prime Intellect \\
\And
  Jack Min Ong \\
  Prime Intellect \\ 
\AND
  Sami Jaghouar \\
  Prime Intellect \\
\And
  Johannes Hagemann \\
  Prime Intellect \\
}

\begin{document}

{
\begingroup
\begin{figure*}
    \centering
    \includegraphics[width=0.125\textwidth]{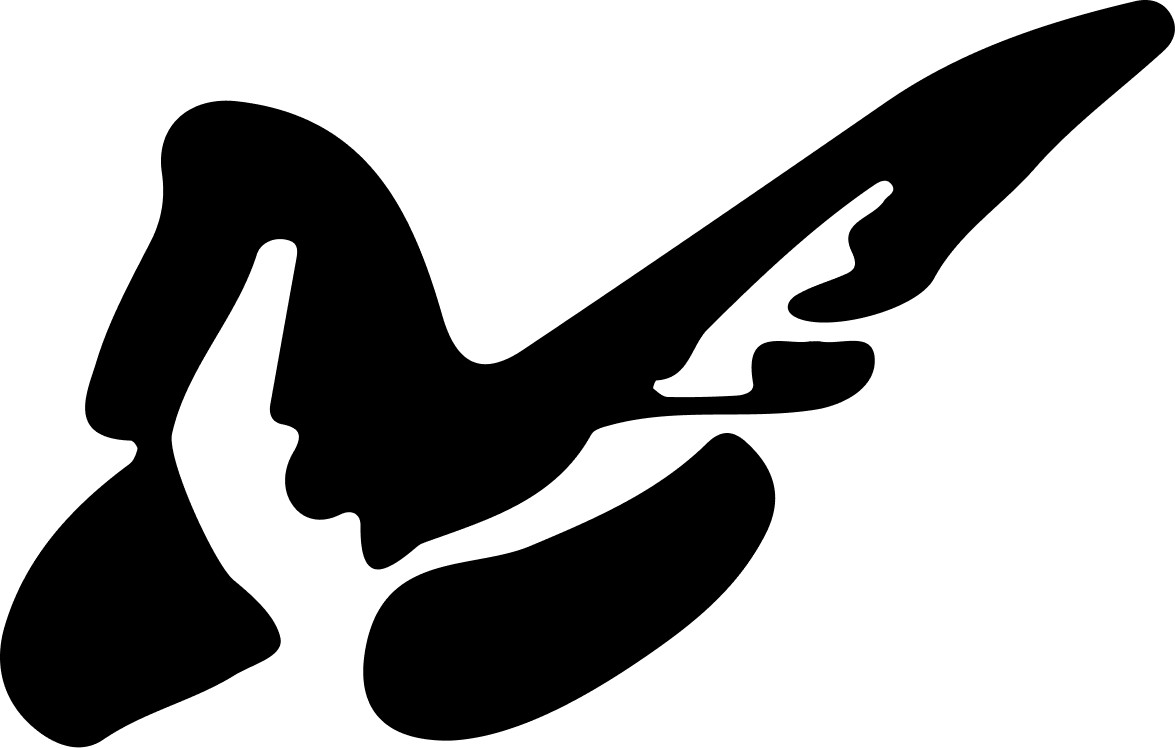}
\end{figure*}
\endgroup
}
\setcounter{figure}{0}

\maketitle
\begin{abstract}
This report presents the Prime Collective Communications Library (PCCL), a novel fault-tolerant collective communication library designed specifically for distributed machine learning workloads over the public internet.
PCCL introduces a new programming model that enables dynamic peer joining and failure recovery.
The library implements efficient collective operations like all-reduce while providing robust fault tolerance mechanisms that allow the system to continue operating even when peers fail or join during ongoing operations.
We demonstrate that PCCL's design enables practical solutions to dynamic membership challenges in workloads with repeated operations and deterministic state advancement.
Our implementation passes extensive stress tests across all major operating systems, showing reliable operation even under rapid peer churn and concurrent collective operations.
By dispatching to multiple connections, we can efficiently utilize cross-continental long-fat-pipe TCP WAN links, in our experiments achieving up to 45 Gbit/s of bandwidth utilization across western Europe and 25 Gbit/s across North America and Europe.
PCCL's architecture enables easy implementation of distributed low-communication optimization strategies like DiLoCo, which significantly reduce communication frequency. Combined with quantization, this leads to a significant reduction in the bandwidth required for distributed training workloads.
PCCL also allows for concurrent collective operations, which enables optimization strategies like async DiLoCo, which can completely hide communication overhead by implementing one-step delayed parameter updates.
PCCL can facilitate exact bit-parity of the shared state across peers in all cases induced by graceful or abrupt peer churn.
While PCCL exposes a C99 API, Python bindings are available which are compatible with PyTorch alongside FSDP.
PCCL is available under the open source MIT license.

\end{abstract}

\newpage

\tableofcontents

\newpage

\section{Introduction}

Over the past decade, the increasing scale and complexity of machine learning (ML) models have driven the adoption of large-scale distributed training across clusters of GPUs or other specialized accelerators. These workloads are especially communication-intensive, as parameter updates and gradient calculations must be synchronized across multiple worker nodes. To handle this synchronization, collective communication libraries like MPI and, more recently, NVIDIA Collective Communications Library (NCCL) \citep{nccl} have played a central role in providing efficient all-reduce, broadcast, and other collective operations on tightly coupled, high-performance clusters.

Despite their success, these traditional libraries were not designed to handle the realities of modern, highly dynamic environments, where resources may join or leave frequently and where connectivity spans public clouds or other wide-area networks. They often assume fixed membership, predictable node connectivity, and uniform performance characteristics—assumptions that rarely hold at large scale or outside dedicated HPC clusters. As a result, when failures or dynamic reconfigurations occur, these libraries do not gracefully handle peer churn, and thus prevent resilient training.

In this work, we introduce the Prime Collective Communications Library (\pccl), a novel fault-tolerant collective communication library that addresses these fundamental limitations. \pccl{} is designed specifically for distributed ML workloads operating over public networks, where peers may fail, join, or disconnect at any point during training. Our approach enables dynamic peer membership and incorporates robust fault-tolerance mechanisms that allow the system to continue collective operations even as peers come and go. By exploiting the deterministic nature of most optimizers, where bit-identical state progression is guaranteed if all peers receive identical updates - even when performed on GPUs, \pccl{} ensures consistent model synchronization across nodes without additional communication beyond the collective operation results.

Unlike traditional HPC libraries, \pccl{} leverages a master-client model that accommodates dynamic membership and continuous topology optimization. New peers join as "registered," then graduate to full participation through a quick consensus phase. Each collective operation proceeds via carefully orchestrated micro-consensus steps—ensuring fault tolerance and consistent model states, even if peers leave mid-operation. In the following sections, we detail how these design decisions enable smooth peer churn, p2p bandwidth measurements for ring-based optimization, and bit-parity among peers inside the shared state without unnecessary re-transmissions.

\section{PCCL Architecture}
\label{sec:pccl-architecture}

\subsection{Overview}
PCCL introduces a novel approach to collective communications for distributed machine learning that differs significantly from classic HPC libraries such as OpenMPI or NCCL.
This section describes the design rationale behind PCCL's master-client model, its fine-grained state machine, and the mechanisms it employs to ensure fault tolerance and dynamic peer membership. 
The following subsections detail each component, providing insights into how PCCL handles joining or failing peers mid-run, optimizes its ring topology based on bandwidth measurements, and enforces bit-identical shared state across participants.

\subsection{Master--Client Model}
\label{sec:master-client-model}
Unlike traditional libraries that assume all processes start simultaneously, PCCL leverages a lightweight master process that orchestrates membership and consensus of collective operations. 
\begin{itemize}
    \item \textbf{Clients (Peers)}: Each training process instantiates a PCCL communicator and connects to the master. Once a peer is "registered," it can be voted into the active group.
    \item \textbf{Master Node}: Maintains a view of which peers are accepted, tracks the ring topology (for all-reduce operations), and coordinates all state transitions through micro-consensus steps.
\end{itemize}

\subsubsection{Membership Lifecycle}
\label{sec:membership-lifecycle}
When a new peer connects, it enters an initial \emph{registered} phase. Accepted peers periodically vote on whether to admit newly registered peers into their group. If they do so, the master instructs everyone to establish direct p2p connections as needed. This approach allows peer churn without forcing restarts of the entire training job.

\subsubsection{What the Master Tracks}
\label{sec:master-tracking}
\paragraph{Clients and Their Status.}
The master differentiates between \emph{registered} peers (not yet active) and \emph{accepted} peers (fully integrated into the collective).

\paragraph{Ring Topology and Bandwidth Metrics.}
Based on p2p bandwidth tests, the master attempts to optimize the ring order, solving an asymmetric traveling-salesman problem \citep{intellect1}.
For this purpose, we have developed a custom heuristic and exact ATSP solver called \emph{libtsp} \citep{libtsp}.
PCCL's optimization can occur as a quick pass (immediate solution within a short timeout) or a more involved "moonshot" background computation. 
Once a new or improved ring order is established, the master instructs peers to reconfigure their p2p connections accordingly.

\paragraph{Shared State.}
The master ensures that each peer has the same shared state.
Any peer with an out-of-date revision or mismatched content is instructed to request data from an up-to-date peer. 
This ensures that all participants converge on a bit-identical set of parameters and optimizer states.

\paragraph{Collective Operations.}
The master coordinates all active collective operations, preventing conflicts between accepting new peers, synchronizing shared state, and performing ring-based collectives.
While it does coordinate the collective operations, it is not involved in the actual collective operations nor shared state re-transmissions themselves.
It merely ensures consensus of all peers that the collective operation has completed and propagates IO failures to potentially interrupt ongoing collective operations to avoid stall.

\subsection{Collective Operations and States}
\label{sec:collective-operations-states}
PCCL manages each operation (e.g., an all-reduce) through well-defined states. 
This state machine approach enables graceful error handling if a peer drops out mid-operation and ensures that every peer transitions in lockstep.

\subsubsection{Phases and State Machine}
\label{sec:phases-state-machine}
Each peer moves through a sequence of phases:
\begin{itemize}
  \item \textbf{ConnectionPhase}: 
    \texttt{PEER\_REGISTERED} or \texttt{PEER\_ACCEPTED}.
  \item \textbf{ConnectionState}: 
    e.g.\ \texttt{IDLE},\\
\texttt{VOTE\_ACCEPT\_NEW\_PEERS}, \texttt{COLLECTIVE\_COMMUNICATIONS\_RUNNING}, 
    etc.
  \item \textbf{CollectiveCommunicationState (per tag)}: 
    \texttt{VOTE\_INITIATE\_COLLECTIVE\_COMMS},\\
    \texttt{PERFORM\_COLLECTIVE\_COMMS}, \texttt{VOTE\_COMPLETE\_COLLECTIVE\_COMMS}.
\end{itemize}

Each transition requires unanimous agreement among involved peers ("micro-consensus") so that inconsistencies or partial failures can be detected and handled quickly.

\subsubsection{One Operation at a Time per Group}
\label{sec:one-operation-per-group}
Within a single peer group, PCCL enforces a strict rule that only one major operation (accepting peers, synchronizing shared state, or running a collective) can be active at any given moment. This ensures correctness and predictability, especially in the face of dynamic membership.

\clearpage

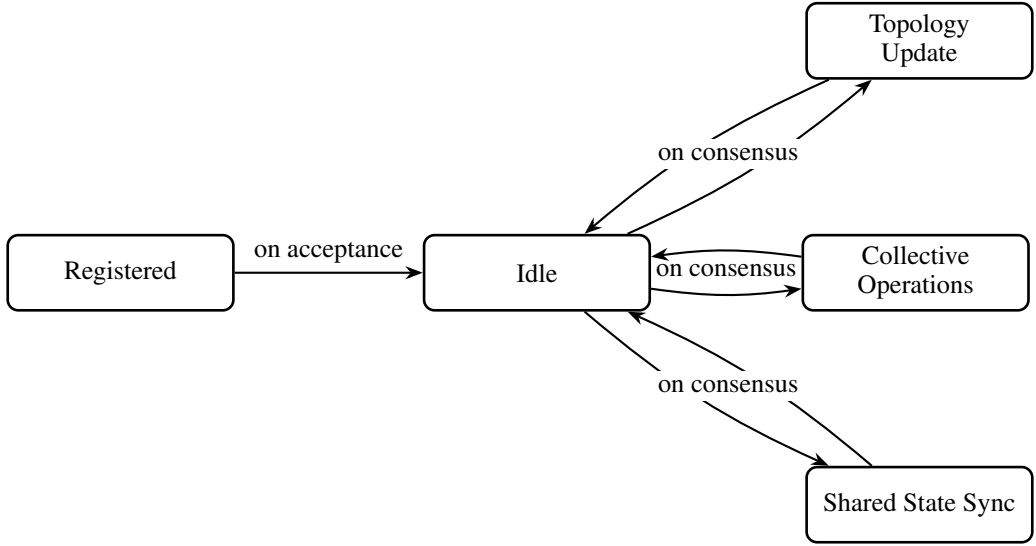
\begin{figure}[ht]
  \centering
  \tikzstyle{state} = [rectangle, 
                       rounded corners, 
                       draw=black, 
                       thick,
                       minimum width=3cm, 
                       minimum height=1cm,
                       align=center]
  \tikzstyle{arrow} = [->, thick]

  \begin{tikzpicture}[
    >=Stealth,
    line width=1pt,
    node distance=2.2cm and 2.5cm,
    state/.style={
      rectangle,
      rounded corners,
      draw=black,
      minimum width=3cm,
      minimum height=1cm,
      align=center
    },
    decision/.style={
      diamond,
      draw=black,
      aspect=2,
      text width=2.2cm,
      align=center,
      inner sep=0pt
    },
    every label/.append style={font=\footnotesize}
]
      \node[state] (registered) {Registered};
      \node[state, right=2.5cm of registered] (accepted) {Idle};
      \node[state, below right=2.9cm of accepted] (sync) {Shared State Sync};
      \node[state, above right=2.9cm of accepted] (optimize) {Topology\\ Update};
      \node[state, right=2.0cm of accepted] (collective) {Collective\\ Operations};

      \draw[arrow] (registered) -- (accepted) node[midway, above, sloped] {on acceptance};

      \draw[arrow] (accepted) to[bend right=8] (sync);
      \draw[arrow] (accepted) to[bend right=8] (collective);
      \draw[arrow] (accepted) to[bend right=8] (optimize);

      \draw[arrow] (sync) to[bend right=8] (accepted);
      \draw[arrow] (collective) to[bend right=8] (accepted);
      \draw[arrow] (optimize) to[bend right=8] (accepted);
      
      \path (accepted) -- (sync) coordinate[midway] (AcAcceptedSync);
      \node[above, fill=white, text=black, yshift=-0.5em] at (AcAcceptedSync) {on consensus};

      \path (accepted) -- (collective) coordinate[midway] (AcAcceptedCollective);
      \node[above, text=black, yshift=-0.5em] at (AcAcceptedCollective) {on consensus};

      \path (accepted) -- (optimize) coordinate[midway] (AcAcceptedOptimize);
      \node[above, fill=white, text=black, yshift=-0.5em] at (AcAcceptedOptimize) {on consensus};

  \end{tikzpicture}
  \caption{Example flow chart for PCCL states and transitions.}
  \label{fig:pccl-flow-chart}
\end{figure}

\subsection{Shared State and Consistency}
\label{sec:shared-state-consistency}
A unique feature of PCCL is its focus on maintaining bit-identical state across all accepted peers. 

Each peer submits a hash of its local parameter (tensor) contents. 
The master identifies any outlier peers that need to be updated, prompting a direct p2p data transfer to bring them into sync.

This process leverages a simple, non-cryptographic, purpose-built, parallelizable hashing function internally referred to as \texttt{simplehash}, which draws inspiration from fvn1a and utilizes a warp-tree reduce \citep{fnv1a}.
This kernel is completely deterministic despite its parallel nature and empirically validated on GPU generations ranging from the GTX 980 Ti to the B200.
To not make assumptions about device placement of shared state across peers, the hash function is also implemented on CPU utilizing OpenMP to emulate the parallel execution model of the GPU while preserving hash identity.
The hash function achieves performance comparable to NVIDIA Thrust's reduce kernels \citep{thrust} \ref{sec:appendix_simplehash_vs_thrust}.

\subsection{Fault Tolerance and Dynamic Membership}
\label{sec:fault-tolerance-membership}
In a realistic setting, peers may fail or disconnect unexpectedly:
\begin{itemize}
    \item If a peer disappears mid-collective, PCCL cancels the current operation, marks the peer as gone, and allows the application developer to retry the collective without that peer.
    \item If the the number of peers drops below a user-specified threshold, training can pause or wait for new arrivals.
\end{itemize}

\subsubsection{Accepting New Peers}
\label{sec:accepting-new-peers}
When a fresh peer connects, it remains in \texttt{PEER\_REGISTERED} until existing accepted peers vote to integrate it. 
At that point, the master instructs everyone to set up p2p connections with the newcomer. 
If any step of that connection process fails, peers can retry or reject the newcomer, preserving stability and consistency.

\subsubsection{Topology Optimization}
\label{sec:topology-optimization}
To perform ring-based collectives efficiently, PCCL measures bandwidth between peers and solves an asymmetric traveling-salesman problem to optimize ring order. 
The master first attempts to find a heuristic solution within a predetermined time limit to avoid stalling the run, while it may later attempt a "moonshot" for a potentially more optimal arrangement.
When a new or improved topology is discovered, PCCL triggers a reconfiguration phase among all accepted peers to adopt the updated ring.
If, for example, computers are co-located in the same datacenter, packets can often be delivered locally without ever bouncing off the gateway.
Thus, topology optimization naturally results in the exploitation of low-latency local communication paths and minimizes the number of expensive public internet hops in the route.

\subsubsection{Fault Tolerance}

Fault tolerance in general is a challenging problem that cannot be solved fully by any single mechanism.
Fault tolerance is a property of the system as a whole that indicates the system's ability to continue operating correctly even in the presence of faults.
If error paths are not unwound correctly, the system could stall, or if the recovery logic is not correct, the system could enter an inconsistent state.

What this amounts to is an engineering problem of identifying all possible error paths.
To make this problem tractable, it is crucial to restrict as much as possible the set of legal state transitions the system can make.

Previous internal attempts at implementing this library have failed to achieve fault tolerance because the API was too permissive.
Peers could join and leave the run any time - as opposed to only via a dedicated accept phase, resulting in a system that created a nearly uncountable number of legal states and error paths.

This justifies the existence of the rather "authoritarian" master node that coordinates micro-consensus steps.
By reducing the system-state to what amounts to a clear finite state machine, we can reason about the system's behavior in a much more straightforward manner.

Ensuring lockstep of application-level state also turned out to be impossible with previous approaches, as join any time means it cannot be known where in the training loop other peers are, thus making it impossible to fall into lockstep.

Only through significantly limiting the set of allowed operations was it possible to achieve a system that is fault tolerant.

\subsection{Low Overhead}
PCCL is best seen as a light-weight wrapper for \texttt{send()} and \texttt{recv()} socket operations with inexpensive error handling and recovery paths.
For a tiny MNIST model with $128$ hidden units (100k parameters), PCCL achieves the following per-api-call latencies over loopback:

\begin{itemize}
  \item \texttt{pcclUpdateTopology} : $0.097667$ ms
  \item \texttt{pcclSynchronizeSharedState} (includes hashing) : $1.1595$ ms
  \item \texttt{pcclAllReduceAsync} : $0.005292$ ms
  \item \texttt{pcclAwaitAsyncReduce} (awaits collective operations): $7.03225$ ms
\end{itemize}

\subsubsection{Stress Testing}
PCCL passes extensive long-running stress tests in CI on all supported operating systems (Linux, macOS, Windows) spanning approximately 8 hours per run.
Each stress-test worker simulates a typical training run involving the typical operations like topology updates, shared-state synchronization, and multiple concurrent collective operations.
The training loop is high-frequency in nature and takes on the order of 100 milliseconds per iteration.
This is in hopes of inducing every possible timing issue over a long time horizon.
Peers are spawned and killed randomly every 500-1000 milliseconds.
As long as the shared state is advanced correctly despite the sustained peer churn, the test is considered passing.
To pass this stress test, PCCL must correctly unwind all possible error paths such that the application can correctly retry in lock-step and that all concurrent operations are scheduled and awaited and retried correctly if a failure occurs in a particular operation.

\section{Implementing Distributed Training Algorithms}

PCCL, while designed for use in distributed training, is a general purpose library for collective communications.
It provides the necessary primitives for implementing distributed computing algorithms.
PCCL does not provide any higher-level distributed training algorithms, but rather provides a foundation on which such algorithms can be built.

We hope that PCCL provides the generality necessary to be useful not only pre-existing distributed training algorithms, but also to be a tool for future research in the design of new distributed training algorithms.

The following sections describe how to implement some popular distributed training algorithms using PCCL:

\subsection{Distributed Data Parallel}
While DDP \citep{Dean2012} is not particularly well-suited for training over the public internet due to its frequent communication requirements, we will start with naive DDP as a starting point for demonstrating the programming paradigm of PCCL.

\begin{algorithm}[H]
  \caption{Naive DDP Training Loop with PCCL (Pseudo-Code)}
  \label{alg:naive-ddp-with-pccl}
  \begin{algorithmic}[1]
  \Require
  \Statex \(\mathcal{C}\) \(\gets\) PCCL communicator
  \Statex \(model\) \(\gets\) local copy of the model
  \Statex \(optimizer\) \(\gets\) local optimizer
  \Statex \(\mathrm{train\_loader}\) \(\gets\) local data loader (each peer sees a distinct subset of the data).

  \Comment{\emph{Collect shared state. (model and optimizer state)}}
  \Statex \(\mathrm{sharedState} \gets [\,p_{data} \;\;|\;\; p_{data} \in model] \cup [\,opt_{state} \;\;|\;\; opt_{state} \in optimizer]\)

  \For{\(iteration = 1\) \textbf{to} \(\mathit{maxIters}\)}

    \Comment{\emph{Update topology and accept new peers}}
    \State \(\mathrm{updateTopology}(\mathcal{C})\)

    \Comment{\emph{Synchronize shared state. (No-op if advanced correctly)}}
    \State \(\mathrm{syncSharedState}(\mathcal{C}, sharedState)\)

    \Comment{\emph{Perform forward and backward pass}}
    \State \(\mathrm{batch} \gets \mathrm{LocalDataLoader.nextBatch()} \)
    \State \(\mathrm{loss} \gets \mathrm{ForwardBackwardPass}(\theta^{(p)}, \mathrm{batch})\)
    \State \(\mathrm{optimizer.zeroGrad()}\)
    \State \(\mathrm{loss.backward()}\)
  
    \Comment{\emph{Use PCCL's All-Reduce to sum gradients across all peers.}}
    \State \(\mathrm{AllReduce}(\mathcal{C}, [\,p.\mathrm{grad} \;\;|\;\; p \in model], \texttt{op} = \textsc{Avg})\)
  
    \Comment{\emph{Perform the local parameter update (identical on each peer).}}
    \State \(\mathrm{optimizer.step()}\)

  \EndFor
  
  \end{algorithmic}
  \end{algorithm}
  
  \noindent
  \textbf{Explanation of Key Steps:}
  \begin{itemize}

    \item \textbf{Synchronize Shared State}:
      The model's parameters and optimizer state are considered shared state.
      The shared state is validated to be identical on all peers and retransmitted if necessary.
      In a correct application loop, this step is a no-op except for the first iteration after a peer has joined the run.

      \item \textbf{Local Forward/Backward Pass}:
      Each peer runs a forward and backward pass on its mini-batch, producing local gradients.
  
      \item \textbf{All-Reduce of Gradients}:
      The PCCL All-Reduce call (\(\mathrm{AllReduce}(\mathcal{C}, \mathcal{G}, \textsc{Avg})\)) averages the local gradients across all peers. After this step, every peer holds the same average of gradients.
   
      \item \textbf{Local Update}:
      Each peer's optimizer applies exactly the same gradient step, so the model parameters remain in sync across all peers.
      In a correct implementation, each peer should independently advance to precisely the same state after this operation.
  \end{itemize}

  \subsection{DiLoCo}

  DiLoCo (\emph{Distributed Low-Communication}) \citep{DiLoCo} is a local-SGD-based optimization strategy inspired by recent work on large-scale distributed training via federated averaging.  It divides the training loop into an \emph{inner} phase and an \emph{outer} phase:
  \begin{itemize}
      \item \textbf{Inner Phase:} Each peer (worker) trains locally for \(H\) steps using a standard optimizer such as AdamW.  Crucially, \emph{no communication} occurs during these inner steps, which drastically reduces overhead over the public internet.
      \item \textbf{Outer Phase:} After \(H\) local steps, each peer computes its parameter difference \(\Delta^{(p)} = \theta - \theta^{(p)}\).  These differences are collectively merged (\emph{e.g.}, averaged) to form an \emph{outer gradient}, which is then applied by an \emph{outer optimizer} (e.g.\ Nesterov momentum).
  \end{itemize}
  Because peers only synchronize once per \(H\) steps, DiLoCo drastically reduces communication required, making it suitable for training over the public internet.

  DiLoCo at number of steps \(H=1\) with an outer optimizer of \texttt{SGD(lr=1.0)} is roughly equivalent to DDP assuming inner optimizer states are roughly comparable across peers by training on similar data.
  DiLoCo can thus be arbitrarily configured on a continuum from what is roughly equivalent to DDP to more infrequent communication at the cost of convergence speed.

  \begin{algorithm}[!htbp]
  \caption{DiLoCo Training Loop (Pseudo-Code)}
  \label{alg:diloco}
  \begin{algorithmic}[1]
  \Require
  \Statex \(\mathcal{C}\) \(\gets\) PCCL communicator
  \Statex \(\{ \theta^{(p)} \}\) \(\gets\) model parameters on each peer \(p\)
  \Statex \(\theta^{(g)}\) \(\gets\) global (outer) aggregator parameters
  \Statex \(\mathrm{optimizer}\) \(\gets\) inner optimizer
  \Statex \(\mathrm{outerOptimizer}\) \(\gets\) outer optimizer
  \Statex \(\mathrm{InnerSteps}\) \(\gets\) number of inner training steps before communication
  \vspace{0.5em}

  \Comment{\emph{Collect shared state. (model and optimizer state)}}
  \Statex \(\mathrm{sharedState} \gets [\,p_{data} \;\;|\;\; p_{data} \in model] \cup [\,opt_{state} \;\;|\;\; opt_{state} \in optimizer]\)
  \linebreak
  
  \For{\(t = 1\) \textbf{to} \(\mathit{maxIterations}\)}

      \State \(\mathrm{updateTopology}(\mathcal{C})\)
      \Comment{\emph{Update topology and accept new peers}}

      \State \(\mathrm{syncSharedState}(\mathcal{C}, \texttt{enforcePopular})\)
      \Comment{\emph{Sync shared state (No-op if advanced correctly)}}

      \Comment{\emph{Local Training Phase}}
      \For{\(i = 1\) \textbf{to} \(\mathrm{InnerSteps}\)}
          \State \(\mathrm{batch} \gets \mathrm{LocalDataLoader.nextBatch()}\)
          \State \(\mathrm{loss} \gets \mathrm{ForwardPass}(\theta^{(p)}, \mathrm{batch})\)
          \State \(\mathrm{optimizer.zeroGrad()}\)
          \State \(\mathrm{loss.backward()}\)
          \State \(\mathrm{optimizer.step()}\)
      \EndFor
   
      \State \(\Delta^{(g)} \gets \theta^{(g)}  - \theta^{(p)}\)
      \Comment{\emph{Aggregate Global Parameter Differences}}
      \State \(\mathrm{AllReduce}(\mathcal{C}, \Delta^{(g)}, \textsc{Avg})\)
      \Comment{\emph{Perform pseudo-gradient reduction}}
  
      \State \(\theta^{(g)} \gets outerOptimizer.step(\theta^{(g)}, \Delta^{(g)})\)
      \Comment{\emph{Apply Global Update}}
      \State \(\theta^{(p)} \gets \theta^{(g)}\)
  
  \EndFor
  
  \end{algorithmic}
  \end{algorithm}
  
  \subsection{Async DiLoCo}
\label{sec:async-diloco}

While DiLoCo reduces communication by only synchronizing after $H$ local steps, \emph{async DiLoCo} goes one step further and \emph{overlaps} the global reduce phase with local compute. This is achieved by maintaining a one-step delay in applying the freshly reduced global update.

For clarity, we note that this is a distinctly different simplified strategy compared to \emph{streaming DiLoCo} \citep{StreamingDiLoCo}.

\paragraph{Key Idea.} Instead of blocking all peers until the all-reduce of parameter deltas finishes, each peer immediately proceeds to the next batch of local training. The background all-reduce completes asynchronously, and its result is incorporated (via the outer optimizer) at the \emph{beginning} of the next outer iteration. As a result, the communication overhead is effectively hidden behind local compute, at the cost of applying parameter deltas one step behind.

The following is a simplified version of Async DiLoCo optimized for clarity.
It assumes that peers do not join or leave the run.
\begin{algorithm}[!htbp]
\caption{Async DiLoCo (Simplified, no peer churn possible)}
\label{alg:async-diloco}
\begin{algorithmic}[1]
\Require
\Statex \quad $\mathcal{C}$ : PCCL communicator
\Statex \quad $\theta^{(p)}$ : local model parameters on peer $p$
\Statex \quad $\theta^{(g)}$ : global (outer) aggregator parameters
\Statex \quad $\mathrm{optimizer}$ : local (inner) optimizer
\Statex \quad $\mathrm{outerOptimizer}$ : outer optimizer acting on $\theta^{(g)}$
\Statex \quad $\mathrm{InnerSteps}$ : \# of local steps before streaming each update
\vspace{3pt}
\State \textbf{Initialize}:
\Statex \quad $\theta^{(p)} \leftarrow \theta^{(g)}$  \Comment all peers start from shared state
\Statex \quad $\Delta^{(\mathrm{t - 1})} \gets \mathbf{0}$ \Comment no pending deltas initially
\Statex \quad $\mathrm{activeThread} \gets \text{None}$ \Comment no active all-reduce thread
\vspace{3pt}

\Comment{\emph{Accept peers in advance, as peer churn is not possible}}
\While {\(\mathrm{getWorldSize}(\mathcal{C}) < \mathrm{targetWorldSize}\)}
\State \(\mathrm{updateTopology}(\mathcal{C})\)
\State \(\mathrm{syncSharedState}(\mathcal{C}, sharedState)\)
\EndWhile

\Comment{\emph{Main Training Loop}}
\For {$t = 1 \dots \mathrm{maxOuterIters}$}
  \For {$i = 1 \dots \mathrm{InnerSteps}$}
    \Comment{\emph{Local Training Phase}}
    \State $(x, y) \gets \mathrm{LocalDataLoader}.\mathrm{nextBatch}()$
    \State $\mathrm{loss} \gets \mathrm{ForwardPass}(\theta^{(p)}, x, y)$
    \State $\mathrm{optimizer}.\mathrm{zeroGrad}()$
    \State $\mathrm{loss}.\mathrm{backward}()$
    \State $\mathrm{optimizer}.\mathrm{step}()$
  \EndFor
  \vspace{2pt}
 
  \Comment{\emph{Await completion of all-reduce of previous iteration's parameter differences}}
  \State \(\mathrm{AwaitAsyncAllReduce}(\mathcal{C}, \Delta^{(t-1)})\)
 
  \State $\Delta^{(t)} \gets \theta^{(g)} - \theta^{(p)}$
  \Comment{\emph{Aggregate Global Parameter Differences}}

  \State $\mathrm{activeThread} \gets \mathrm{AsyncAllReduce}(\mathcal{C}, \Delta^{(t)})$
  \Comment{\emph{Launch async all-reduce in background}}
  \vspace{2pt}

  \If {$\mathrm{isFinished}(\Delta^{(t - 1)})$}
    \State $\mathrm{outerOpt}.\text{step}(\theta^{(g)},\;\Delta^{(\mathrm{t - 1})})$
    \Comment{\emph{Apply global update of previous iteration}}
    \State $\theta^{(p)} \gets \theta^{(g)}$
  \EndIf

  \State ~\textbf{Proceed immediately to next iteration; 
         all-reduce overlaps with local compute.}
\EndFor
\end{algorithmic}
\end{algorithm}

\paragraph{Implementation Details.} 
\begin{itemize}
    \item Each peer keeps a \emph{background thread} or analogous asynchronous mechanism to run \texttt{all-reduce} on the \emph{previous} iteration's \(\Delta^{(t - 1)}\). Thus, no peer stalls waiting for communication.
    \item At iteration \( t \), if the previous all-reduce is complete, we apply the newly aggregated result via the \(\mathrm{outerOptimizer}\). Then local training for iteration \( t \) proceeds normally.
    \item At the \emph{end} of iteration \( t \), we compute $\Delta^{(t)}$ and immediately hand it off for asynchronous all-reduce. Meanwhile, iteration \( t{+}1 \) starts. 
\end{itemize}

In practice, \emph{async} DiLoCo must also handle situations where a new peer joins (or an existing peer fails) during training. 
Because PCCL requires that major operations (topology updates, shared-state syncs, or collectives) do \emph{not} overlap with collective operations, 
we periodically check whether there are \emph{pending} peers before awaiting the completion of any in-flight collective to start e.g a topology update and subsequent shared-state synchronization.
If there are new arrivals, we safely pause any in-flight collective, update topology, and run a shared-state synchronization 
exactly once, and then immediately resume async DiLoCo.

\subsubsection{Handling Peer Churn in Async DiLoCo}
\label{sec:async-diloco-churn-explained}

While Async DiLoCo (see \Cref{alg:async-diloco}) effectively overlaps local compute with background communication, additional considerations arise when peers join or leave mid-run. In particular, each collective operation (the all-reduce of parameter deltas) is spread across multiple training iterations because of the one-step delay. Therefore, any membership change must be carefully sequenced to avoid interfering with in-flight collectives, which PCCL does not allow to overlap for safety reasons.

\paragraph{1.\ No Overlap of Major Operations with Collectives.}
As described in \Cref{sec:pccl-architecture}, PCCL restricts a communicator to only one major operation at a time:
\begin{itemize}
    \item accepting or removing peers (topology updates),
    \item collective communication (e.g.\ all-reduce),
    \item shared state synchronization.
\end{itemize}
A new peer that arrives is placed in the \emph{registered} state until the pre-existing set of peers accept the new peer. Similarly, if a peer fails in the middle of a collective, PCCL aborts the operation, reconfigures membership (without accepting new peers), and only then allows the next collective to begin.

\paragraph{2.\ Reconciling One-Step-Behind Updates.}
In \Cref{alg:async-diloco}, every peer applies global parameter deltas with a one-iteration delay so local training can run in parallel with background communication. When a new peer arrives, it must:
\begin{enumerate}
    \item \textbf{Synchronize Shared State:} Obtain the most recent aggregator (outer) parameters and optimizer state from an active peer.
    \item \textbf{Reset Inner Model:} Copy those aggregator parameters into its local model, ensuring it starts from the current global baseline.
    \item \textbf{Enter the One-Step-Behind Pipeline:} On the next iteration, the new peer contributes its parameter deltas in the background all-reduce. Because it "misses" the completion of the all-reduce of the previous iteration, additional synchronization is required for the newcomer to enter into lock-step with bit-parity.
\end{enumerate}

\paragraph{3.\ Temporarily stalling the pipeline for membership changes.}
As mentioned in section \ref{sec:pccl-architecture}, PCCL requires that major operations (topology updates, shared-state syncs, or collectives) do \emph{not} overlap with collective operations.
This means that when a new peer arrives, we must temporarily stall the pipeline until the in-flight collectives complete.
Only then can we proceed with the membership change and the subsequent shared-state synchronization.
PCCL thus provides a mechanism to check for pending peers to only trigger the above steps if there are pending peers
that require integration.
This function is guaranteed to return the same value for all peers in the run.
This function employs a micro-consensus scheme similar to the major operations, however, it can be arbitrarily overlapped with concurrent collectives.

\paragraph{4.\ Ensuring Global Consistency.}
Once a new peer is accepted, the application triggers a shared-state synchronization.
This brings the peer up to date with the latest global state. The inner steps of the new peers will
be based on the same global state as those of the other peers.
The pseudo-grad output of the resulting inner steps is thus "weight-compatible" with that of the other peers
and the changes from the current outer step can be contributed into the next all-reduce.
However, because the new peer joined after the previous all-reduce has already started,
which will determine the \emph{next} outer state, the new peer must "eavesdrop" on the result of the
previous all-reduce to acquire the correct outer state.
Thus, one additional shared state synchronization is required in the outer step when a new peer joins.
The pre-existing peers provide the shared state after being advanced by the result of the previous all-reduce
that the new peer missed.
The new peer simply adopts the outer state provided by the pre-existing peers without performing an outer optimizer step.
After this procedure is complete, the new peer is now fully integrated and "weight-compatible" with the other peers.

The second shared state synchronization utilizes the \texttt{sendOnly} or \texttt{receiveOnly} strategies to ensure that the shared state of newcomers is never chosen to be distributed.
This is important, for example, if more peers join in one step than the number of pre-existing peers.
If this is the case and we would use the default \texttt{enforcePopular} strategy, the pre-existing peers will loose hash popularity and the undesired state is chosen and distributed.

Allowing peer churn while sustaining a \emph{one-step-behind} global update enables robust scaling in unreliable or heterogeneous environments.
DiLoCo's inner steps $H$ can be set such that $t_{compute}$ roughly equals $t_{comm}$ for a balanced system.
In such a setting, the communication overhead is completely hidden behind local compute.
Future work may explore under which conditions waiting for communication "amortizes" in terms of better wall-clock convergence.

\pagebreak

\begin{algorithm}[htbp]
  \caption{Async DiLoCo with Conditional Topology \& Shared State Updates}
  \label{alg:async-diloco-churn}
  \begin{algorithmic}[1]
  \Require
    \Statex \quad $\mathcal{C}$ \hfill \text{PCCL communicator}
    \Statex \quad $\theta^{(g)}$ \hfill \text{Aggregator (outer) parameters (shared state)}
    \Statex \quad $\theta^{(p)}$ \hfill \text{Local model parameters on peer $p$}
    \Statex \quad $\mathrm{optimizer}$ \hfill \text{Local inner optimizer}
    \Statex \quad $\mathrm{outerOpt}$ \hfill \text{Outer optimizer acting on $\theta^{(g)}$}
    \Statex \quad $\mathrm{InnerSteps}$ \hfill \text{Number of local steps per iteration}
    \Statex \quad $\mathrm{activeThread} \gets \text{None}$ \hfill \text{No initial all-reduce thread}
    \Statex \quad $\Delta^{(\mathrm{t - 1})} \gets \mathbf{0}$ \hfill \text{No pending delta initially}
  \Statex
  
  \For{$t = 1 \dots \mathrm{maxOuterIters}$}
    \State \textbf{Check for new peers (only if needed)}
    \If{$\mathcal{C}.\texttt{are\_peers\_pending}()$}
       \Comment{Handle membership changes now}
       \If{$\mathrm{activeThread}$ still running}
       \State \(\mathrm{AwaitAsyncAllReduce}(\mathcal{C}, \Delta^{(t-1)})\)
         \Comment wait for any in-flight collective
       \EndIf
       \State $\mathcal{C}.\texttt{updateTopology}()$
       \Comment Accept new peers
       \State $\mathcal{C}.\texttt{syncSharedState}(\theta^{(g)}, \texttt{enforcePopular})$ 
       \Comment e.g.\ aggregator \& outerOpt state
       \State $\theta^{(p)} \gets \theta^{(g)}$ 
       \Comment local model is now in sync
    \EndIf
    \medskip
    \Comment{\emph{Local (inner) training}}
    \For {$i = 1 \dots \mathrm{InnerSteps}$}
      \State $(x, y) \gets \mathrm{LocalData.nextBatch}()$
      \State $\mathrm{optimizer}.\text{zeroGrad}()$
      \State $\mathrm{loss} \gets \mathrm{ForwardBackwardPass}(\theta^{(p)}, x, y)$
      \State $\mathrm{optimizer}.\text{step}()$
    \EndFor
    
    \vspace{2pt}
    \State \(\mathrm{AwaitAsyncAllReduce}(\mathcal{C}, \Delta^{(t-1)})\)
    \Comment wait for any in-flight all-reduce

    \State $\Delta^{(t)} \gets \theta^{(g)} - \theta^{(p)}$
    \Comment{\emph{Compute new parameter difference}}

    \State $\mathrm{activeThread} \gets \mathrm{AsyncAllReduce}(\mathcal{C},\;\Delta^{(t)})$
    \Comment{\emph{Launch async all-reduce in background}}

    \If {$\mathrm{isFinished}(\Delta^{(t - 1)})$}
      \State $\mathrm{outerOpt}.\text{step}(\theta^{(g)},\;\Delta^{(\mathrm{t - 1})})$
      \Comment{\emph{Apply global update of previous iteration}}
      \State $\theta^{(p)} \gets \theta^{(g)}$
      \If {$\mathrm{newComerHasJoined}$}
        \State $\mathcal{C}.\texttt{syncSharedState}(\theta^{(g)}, \texttt{sendOnly})$ 
        \Comment{\emph{Provide result of previous all-reduce}}
      \EndIf
    \Else
      \If {$\mathrm{selfIsNew}$}
        \State $\mathcal{C}.\texttt{syncSharedState}(\theta^{(g)}, \texttt{receiveOnly})$
        \Comment{\emph{Obtain result of previous all-reduce}}
      \EndIf
      \State $\theta^{(p)} \gets \theta^{(g)}$
    \EndIf
    \EndFor
  \end{algorithmic}
\end{algorithm}
\pagebreak

\section{Implementation Details}
\subsection{All-Reduce Implementation}

Below is pseudo-code for the ring-based All-Reduce operation in PCCL. 
PCCL implements an $N - 1$ formulation of a pipelined ring-reduce, where $N$ represents the world-size.

\begin{algorithm}[!htbp]
\caption{Ring-Based All-Reduce (High-Level Pseudo-Code)}
\label{alg:ring-allreduce}
\begin{algorithmic}[1]
\Statex \textbf{Input:}
\Statex \quad \texttt{buffer}: Local array/tensor of size $N$
\Statex \quad \texttt{data\_type}: Element type (e.g., \texttt{float32})
\Statex \quad \texttt{reduce\_op}: Reduction operator (e.g.\ \textsc{Sum})
\Statex \quad \texttt{rank}: This peer's index in $[0..\text{world\_size}-1]$
\Statex \quad \texttt{world\_size}: Total number of peers
\Statex \quad \texttt{ring\_order}: Order of ranks in the ring
\vspace{2mm}
\Statex \textbf{Output:}
\Statex \quad \texttt{buffer}: Overwritten so that all peers share the fully reduced result
\vspace{1mm}
\State $N \gets \text{length}(\texttt{buffer})$
\State \texttt{chunkBoundaries} $\gets \textproc{ComputeChunkBoundaries}(N, \texttt{world\_size})$
\Comment{$\texttt{chunkBoundaries}[r] = (start\_idx, end\_idx)$}

\State $(\texttt{selfStart}, \texttt{selfEnd}) \gets \texttt{chunkBoundaries}[\texttt{rank}]$

\State $(\texttt{buffer\_bak}) \gets \texttt{clone(buffer)}$

\vspace{1mm}
\Comment{\textbf{Phase 1: Reduce-Scatter}}
\For{$\texttt{step} \gets 0$ \textbf{to} $(\texttt{world\_size} - 2)$}
   \State $\texttt{txChunk} \gets (\texttt{rank} - \texttt{step}) \bmod \texttt{world\_size}$
   \State $\texttt{rxChunk} \gets (\texttt{rank} - \texttt{step} - 1) \bmod \texttt{world\_size}$

   \State $(txStart, txEnd) \gets \texttt{chunkBoundaries}[\texttt{txChunk}]$
   \State $(rxStart, rxEnd) \gets \texttt{chunkBoundaries}[\texttt{rxChunk}]$

   \State $\texttt{txSpan} \gets \texttt{buffer}[txStart : txEnd]$
   \State $\texttt{rxSpan} \gets \texttt{buffer}[rxStart : rxEnd]$

   \State $\textproc{runReduceStage}(\texttt{txSpan}, \texttt{rxSpan}, \texttt{reduce\_op}, \dots)$
   \Comment{Sends txSpan, receives another chunk, accumulates into rxSpan}
   \If{\textit{operation aborted or I/O failure}}
      \State $\texttt{buffer} \gets \texttt{buffer\_bak}$
      \Comment{Restore buffer to pre-operation state}
      \State \textproc{HandleAbortAndReturn}()
   \EndIf
\EndFor

\vspace{1mm}
\Comment{\textbf{Phase 2: Reduce-Gather}}
\State $\texttt{currentChunk} \gets (\texttt{rank} + 1) \bmod \texttt{world\_size}$

\For{$\texttt{step} \gets 0$ \textbf{to} $(\texttt{world\_size} - 2)$}
   \State $(txStart, txEnd) \gets \texttt{chunkBoundaries}[\texttt{currentChunk}]$
   \State $\texttt{txSpan} \gets \texttt{buffer}[txStart : txEnd]$

   \State $\texttt{incChunk} \gets (\texttt{currentChunk} - 1 + \texttt{world\_size}) \bmod \texttt{world\_size}$
   \State $(rxStart, rxEnd) \gets \texttt{chunkBoundaries}[\texttt{incChunk}]$
   \State $\texttt{rxSpan} \gets \texttt{buffer}[rxStart : rxEnd]$

   \State $\textproc{runAllgatherStage}(\texttt{txSpan}, \texttt{rxSpan}, \dots)$
   \Comment{Broadcast local chunk, receive next chunk into rxSpan}
   \If{\textit{operation aborted or I/O failure}}
      \State $\texttt{buffer} \gets \texttt{buffer\_bak}$
      \Comment{Restore buffer to pre-operation state}
      \State \textproc{HandleAbortAndReturn}()
   \EndIf

   \State $\texttt{currentChunk} \gets \texttt{incChunk}$   \Comment{Take ownership of newly received chunk for next step}
\EndFor

\vspace{1mm}
\Comment{\textbf{Finalize, e.g.\ apply averaging if reduce\_op = \textsc{Avg}}}
\State $\textproc{finalizeReduction}(\texttt{buffer}, \texttt{reduce\_op}, \texttt{world\_size})$
\end{algorithmic}
\end{algorithm}

\begin{algorithm}[!htbp]
  \caption{runReduceStage/runAllgatherStage (structural view)}
  \label{alg:run-reduce-stage-struct}
  \begin{algorithmic}[1]
  \Function{runReduceStage/runAllgatherStage}{%
    \texttt{txSpan},   \Comment{data to send}

    \texttt{rxSpan},   \Comment{where to accumulate}
    
    \texttt{recvBuf},  \Comment{temporary receive buffer}
    
    \texttt{tag}, \texttt{seqNr}, 
    \texttt{op}, 
    \texttt{quantEnabled}, 
    \texttt{rank}, \texttt{worldSize}, \texttt{ringOrder}, 
    \texttt{masterSocket}, 
    \texttt{peerTx}, \texttt{peerRx}}
    
    \State \Comment{Identify next/prev peers in the ring}
    \State $\mathit{txPeer}\gets(\mathit{rank}+1)\bmod \mathit{worldSize}$
    \State $\mathit{rxPeer}\gets(\mathit{rank}-1+\mathit{worldSize})\bmod \mathit{worldSize}$
    \vspace{1mm}
    
    \If{\texttt{quantEnabled}}
      \Comment{Exchange quantization metadata if enabled}
      \State sendMeta(\texttt{peerTx},\,\texttt{tag},\,\texttt{seqNr})
      \State $\textit{meta}\gets receiveMeta(\texttt{peerRx},\,\texttt{tag},\,\texttt{seqNr})$
      \If{abort signaled via \texttt{masterSocket}}
        \State \Return \texttt{(fail, aborted)}
      \EndIf
    \EndIf

    \vspace{1mm}
    \Comment{Full-duplex chunked send/recv loop}
    \State $\mathit{doneSend}\gets\texttt{false},\;\mathit{doneRecv}\gets\texttt{false}$
    \While{$\neg(\mathit{doneSend}\,\wedge\,\mathit{doneRecv})$}
      \State try\_send\_next\_chunk(\texttt{peerTx},\,\texttt{txSpan},\,\texttt{seqNr})
      \State try\_recv\_next\_chunk(\texttt{peerRx},\,\texttt{recvBuf},\,\texttt{seqNr})
      \State if new full chunk arrived \textbf{then}
        \State \quad reduce\_or\_write\_into(\texttt{rxSpan},\,\texttt{recvBuf},\,\texttt{op})
      \State $\texttt{abortReceived}\gets\texttt{masterSocket.recvQueue}.pop()$
      
      \Comment{Periodically check \texttt{masterSocket} for abort (no IO overhead)}
      
      \If{\texttt{abortReceived}}
        \State \Return \texttt{(success, aborted)}
      \EndIf
    \EndWhile
    \vspace{1mm}
    \State \Return \texttt{(success, not\_aborted)}
  \EndFunction
  \end{algorithmic}
  \end{algorithm}

\paragraph{Overview of Steps.}
\begin{enumerate}
    \item \textbf{Compute Chunk Boundaries.} Partition the tensor into $\texttt{world\_size}$ slices. 
    \item \textbf{Reduce-Scatter Phase.} Each rank progressively sends (and receives) slices and accumulates the incoming data into its local buffer using the chosen \texttt{reduce\_op}. After $\texttt{world\_size}-1$ iterations, each rank "owns" exactly one fully reduced chunk.
    \item \textbf{Reduce-Gather Phase.} Each rank broadcasts its reduced chunk around the ring until all peers obtain the entire reduced tensor.
    \item \textbf{Finalize Reduction.} If the requested operation is \textsc{Avg}, each element in the final buffer is divided by $\texttt{world\_size}$. For \textsc{Sum}, \textsc{Max}, etc.\ this step is either a no-op or a simple transform.
\end{enumerate}

Conceptually, it consists of:
\begin{enumerate}
    \item \emph{Reduce-Scatter Phase:} each rank accumulates the partial sums for one slice of the tensor.
    \item \emph{Reduce-Gather Phase:} each rank then broadcasts its completed slice so that all ranks share the final result.
\end{enumerate}

If at any point an IO failure occurs, the operation is aborted and unwinds and restores the buffer to its pre-operation state.

During the reduce-scatter phase, each rank sends and receives a subset of the data to reduce while periodically checking the \texttt{masterSocket} for abort signals.
It is crucial that the abort checking does not induce any additional IO overhead and the corresponding IO operations are performed asynchronously.
We can achieve this check to be largely free as the master connection is a separate TCP stream and the abort signal is pushed from a separate thread.

\section{Implementation Challenges}

\subsection{Socket Implementation Behavior}
PCCL targets TCP/IP for use over the public internet. 
OS-level socket APIs however often differ drastically between operating systems.
Socket header compatibility can often be non-existent, requiring different error handling (e.g on WSA, which
differs drastically from more traditional BSD-style socket implementations) or workarounds for lack of features.
Beyond header incompatibility, socket implementations can often differ drastically in behavior, making it challenging to make tight assertions on behavior within the bounds of the implementation's stated and soft guarantees. This is especially true with respect to blocking/unblocking behavior of functions like \texttt{recv()} in response to \texttt{close()} and \texttt{shutdown()} calls,
half-close drain behavior differing on Windows, and other subtle differences.
We thus relied on extensive CI testing and long-running stress testing simulating extreme conditions of peer churn to validate behavior and to guide design decisions.
During the course of development, we have observed numerous quirks about socket implementation behavior in containerization platforms like Docker
and in some cases even what we believe to be bugs in the XNU kernel with respect to send buffer exhaustion handling.

\subsection{Threadpark}
To saturate full-duplex links, dedicated send and receive threads are used per connection.
However, C++ synchronization primitives such as condition variables proved insufficient in terms of wake-up latency.
We thus implemented a custom library \emph{threadpark} which allows easy implementation of low-latency wake-up systems and lock-free data-structures.
To ensure high-throughput and low-latency, \emph{threadpark} uses futex-like apis on all major operating systems (Linux, macOS, Windows, FreeBSD \& OpenBSD).
This includes Linux's \texttt{futex}, macOS's \texttt{\_\_ulock\_wait} and Windows's \texttt{WakeByAddressSingle}.

\subsection{Zero-Copy, Bit-wise deterministic, Interruptable, Pipelined, Quantized, Asynchronous \& Concurrent Collective Commmunications}
PCCL implements zero-copy, interruptable pipelined collective communications operations, where user-provided buffers (i.e PyTorch tensor storage buffers) are referenced directly by the resulting \texttt{send()} socket operations.

As PCCL must unwind correctly from any IO failure without leaving the system in an inconsistent state or causing undefined behavior,
special care must be taken with respect to
\begin{itemize}
  \item Propagating via explicit means localized failure conditions such that all peers can react
  \item Discarding partial data attributed to aborted operations with minimal protocol state
  \item Phrasing the main algorithm such to allow for concurrent polling of state changes propagated via the master without performance degradation
\end{itemize}

The multi-threaded nature of the algorithm paired with a strictly necessary zero-malloc policy for the reduce code path
necessitated a custom caching allocator to provide the conceptual benefits of dynamic memory allocation without the performance penalties associated with the use of standard \texttt{malloc()} and \texttt{free()} calls and page faults caused by lazy initialization of pages by a first-touch \texttt{memcpy()} call.

\subsubsection{Bit-wise Determinism}
It is well known that a pipelined ring-reduce produces bit-wise deterministic results on all peers given that the share phase propagates the owned subsets of respective peers to all other peers without modification.

However, for most quantization functions that are used in ML training $D(Q(x)) \neq x$, meaning that the quantization function $Q(x)$ and the dequantization function $D(x)$ are not perfect inverses to each other.

Additionally, each peer has access to higher precision data of the subset of data that is its contribution.
For the sake of bit-wise determinism, we cannot make use of the additional precision beyond what can be recovered from dequantization of the data that a peer sends.
Not doing so would result in "lingering" precision present in the fully reduced result of a peer's contribution that is not present on the respective other peers.
However, because accumulation takes place in higher precision than the received quantized data, the argument can be made that as world size grows, we get a better approximation of the true sum and it is thus not necessary to cling to this precision.

\subsection{Determinism of NVIDIA PTX Intrinsics}
To facilitate deterministic advancement of the shared state based on collective operations, it is important to be aware of the kernels invoked in the outer optimizer step.
If the implementation of numerics differs between peers due to different hardware, advancement of the shared state would be non-deterministic.
GPUs are known to be non-deterministic due to their parallel nature, but as long as no true GPU indeterminism is introduced when memory is shared across threads in an inherently race-prone manner,
each thread can be made to execute the same code and produce the same result.

NVIDIA provides a sometimes broken, soft guarantee of forward compatibility of PTX code across different generations of hardware - however not necessarily that
those instructions will behave identically across different generations of hardware.
With the exception of some early PTX 1.1 ISA removals (e.g. \texttt{cross}, \texttt{dot}, \texttt{mag} and \texttt{vred}) \citep{ptx_isa_1_4} -
which were never emitted by any compiler - and the removal of no-sync \texttt{shfl}/\texttt{vote} instructions in PTX 6.4 \citep{ptx_isa_6_4},
most PTX code is generally forward-assemblable.
Additionally, NVIDIA's newly introduced \texttt{sm\_XXXa} compilation target specifically targets architecture-specific features and \texttt{sm\_XXXf} targets features that are family-specific \citep{ptx_isa_8_0}.
Additionally, some intrinsics like \texttt{\_\_nv\_expf} are not real PTX instructions, but are implemented in \texttt{libdevice}, which is an LLVM bitcode module.
Here behavior could either change through changes in \texttt{libdevice} or changes in the ISA implementation across different architectures.
Despite this increased surface area, we confirm \texttt{\_\_nv\_expf} behaves identically on \texttt{sm\_52} (GTX 980 Ti), \texttt{sm\_61} (GTX 1060), \texttt{sm\_89} (RTX 4090), \texttt{sm\_90} (GH 200), \texttt{sm\_100} (B200), as determined by hashing the output bits of \texttt{\_\_nv\_expf} across all 2$^{32}$ bit-patterns of float.
\ref{sec:appendix_expf_determinism}

Due to this result, we can be confident in a defacto guarantee of cross-architecture bit-wise determinism for the \texttt{\_\_nv\_expf} intrinsic and likely many other NVIDIA libdevice/PTX intrinsics.
\clearpage

\section{Benchmarks}

\subsection{HPC Benchmark}
The following benchmark compares the implementation agnostic reduce throughput of PCCL to that of Gloo \citep{Gloo}, as computed by the size of the reduce contribution of one peer divided by the time it takes to reduce across all ranks.
Note that this value is not a measurement of actual communication performed, but rather a way to compare the relative performance of the two libraries in an implementation agnostic manner.
While PCCL does not specifically specialize for high-speed networks, PCCL remains competitive with Gloo.
It should be noted that PCCL's abortable design places PCCL at a disadvantage compared to Gloo, as PCCL must perform many futex syscalls during the course of performing a collective operation.
Thus PCCL is more sensitive to OS scheduler-induced latencies and achieved throughput is therefore subject to higher variance.
Future versions of PCCL may address these issues.

\begin{flushleft}
  \centering
  \begin{tikzpicture}
    \begin{axis}[
        width=10cm,
        height=6cm,
        xlabel={World Size},
        ylabel={Reduce Throughput (MB/s)},
        legend pos=north east,
        xtick={2,3,4,5,6,7,8},
        ymin=0,
        ymajorgrids=true,
        grid style=dashed
      ]
  
      \addplot[
        color=blue,
        thick,
        mark=o
      ] coordinates {
        (2,1455.72)
        (3,1247.40)
        (4,1203.1)
        (5,1099.77)
        (6,1060.78)
        (7,1049.90)
        (8,1019.25)
      };
      \addlegendentry{Gloo}
  
      \addplot[
        color=red,
        thick,
        mark=square
      ] coordinates {
        (2,2027.76)
        (3,1452.74)
        (4,1134.17)
        (5,1012.74)
        (6,948.06)
        (7,999.25)
        (8,958.78)
      };
      \addlegendentry{PCCL}
  
    \end{axis}
  \end{tikzpicture}
\end{flushleft}
The experiment was conducted on eight nodes with approximately 35.6 GBit/s of achievable Ethernet bandwidth over a single connection while reducing $1024\cdot1024\cdot256\cdot4$ = 1.073 GB of data per peer.

\subsection{Effect of Multiple Connections over WAN}

On the public Internet, opening multiple simultaneous connections can dramatically boost achievable bandwidth. TCP's receive-window auto-scaling on high-latency, high-bandwidth WAN links rarely reaches the peak speed that brute-forcing a single connection can achieve—and disabling auto-scaling altogether almost always makes performance worse.
Moreover, many routers use per-flow fair-queuing. By running several parallel streams, one can effectively “carve out” extra bandwidth slices that a single flow alone wouldn't receive under strict per-flow fairness.
Multiple concurrent all reduces can thus dispatch to multiple concurrent connections of the PCCL p2p connection pool, resulting in a higher bandwidth utilization.
This assumption holds for modern ML training given that multiple parameter tensors can be reduced independently,
or potentially even chunked beyond that for element-wise reduction operations.
When performing multiple concurrent all reduces, the effective reduce throughput is computed as the sum of the sizes of the reduce contributions of a particular peer divided by the time it takes for all all reduces to complete.
\pagebreak

As an example, when measuring the total throughput (RX+TX) as a function of number of connections with \texttt{iperf3}, we observe the following scaling from \texttt{europe-west1-b} to \texttt{europe-west2-b}:

\begin{flushleft}
\centering
\begin{tikzpicture}
  \centering
  \begin{axis}[
    xlabel={Number of Concurrent Connections},
    ylabel={Throughput GBit/s (RX+TX)},
    grid=major,
    width=12cm,
    height=6cm,
    ymin=0,
  ]
    \addplot[
      thick
    ] coordinates {
      (1,5.96)   (2,6.12)   (3,12.22)  (4,14.90)  (5,17.36)
      (6,17.80)  (7,20.60)  (8,21.60)  (9,25.80)  (10,31.00)
      (11,34.20) (12,35.40) (13,35.60) (14,38.40) (15,38.40)
      (16,39.60) (17,43.60) (18,45.00) (19,47.00) (20,48.40)
      (21,50.60) (22,51.20) (23,50.60) (24,55.20) (25,54.60)
      (26,54.60) (27,53.20) (28,56.80) (29,57.80) (30,57.40)
      (31,55.40) (32,56.00) (33,56.80) (34,56.20) (35,52.20)
    };
  \end{axis}
\end{tikzpicture}
\end{flushleft}

This effect is even more pronounced for cross-continental connections, as shown in the following figure:
\begin{flushleft}
\centering  
\begin{tikzpicture}
  \centering
  \begin{axis}[
    xlabel={Number of Concurrent Connections},
    ylabel={Throughput (Gbit/s, TX+RX)},
    grid=major,
    width=12cm,
    height=6cm,
    ymin=0,
  ]
    \addplot[
      thick
    ] coordinates {
      (1,0.57)   (2,1.142)  (3,1.706)  (4,2.24)
      (5,2.86)   (6,3.34)   (7,3.98)   (8,4.52)
      (9,5.10)   (10,5.64)  (11,6.24)  (12,6.76)
      (13,7.38)  (14,7.94)  (15,8.52)  (16,9.06)
      (17,9.62)  (18,10.22) (19,10.74) (20,11.30)
      (21,11.92) (22,12.42) (23,13.02) (24,13.58)
      (25,14.14) (26,14.66) (27,15.26) (28,15.84)
      (29,16.38) (30,16.98) (31,17.54) (32,18.02)
      (33,18.62) (34,19.20) (35,19.76) (36,20.40)
      (37,20.80) (38,21.40) (39,22.00) (40,22.40)
      (41,23.20) (42,23.60) (43,24.20) (44,24.80)
      (45,25.20) (46,25.80) (47,26.20) (48,26.80)
      (49,27.60) (50,28.00) (51,28.40) (52,29.20)
      (53,29.60) (54,30.20) (55,30.80) (56,31.40)
      (57,32.00) (58,32.40) (59,32.80) (60,33.40)
      (61,34.00) (62,34.60) (63,35.00) (64,35.60)
      (65,36.20) (66,36.80) (67,37.20) (68,37.60)
      (69,38.20) (70,38.80) (71,39.40) (72,39.80)
      (73,40.40) (74,40.80) (75,41.40) (76,42.00)
      (77,42.60) (78,43.20) (79,43.60) (80,44.20)
      (81,44.60) (82,45.20) (83,45.60) (84,46.40)
      (85,46.80) (86,47.20) (87,47.80) (88,48.40)
      (89,48.80) (90,49.60) (91,49.60) (92,49.80)
      (93,50.00) (94,50.60) (95,50.80) (96,51.00)
      (97,52.60) (98,53.40) (99,53.40) (100,53.00)
      (101,53.00)(102,52.80)(103,53.80)(104,50.40)
      (105,49.20)(106,49.20)(107,48.80)(108,49.40)
      (109,50.60)(110,49.60)
    };
  \end{axis}
\end{tikzpicture}
\end{flushleft}

This behavior is not exclusive to Google Cloud's \texttt{Tier\_1} networking tier, but can also be observed with many other providers.
For example, when measuring the throughput from Nebius (Helsinki, Finland) to Voltage Park (Dallas, Texas) with \texttt{iperf3}, we observe the following scaling:

\begin{flushleft}
\centering  
\begin{tikzpicture}
  \centering
  \begin{axis}[
    xlabel={Number of Concurrent Connections},
    ylabel={Throughput (Gbit/s, TX+RX)},
    grid=major,
    width=12cm,
    height=6cm,
    ymin=0,
  ]
    \addplot[
      thick
    ] coordinates {
      (1,0.592)   (2,0.958)   (3,1.260)  (4,1.562)  (5,1.884)
      (6,2.120)   (7,2.500)   (8,2.680)  (9,2.880)  (10,3.480)
      (11,3.700)  (12,3.980)  (13,4.260) (14,4.520) (15,4.840)
      (16,5.060)  (17,5.600)  (18,5.800) (19,5.800) (20,6.660)
      (21,6.720)  (22,7.240)  (23,7.300) (24,7.700) (25,8.060)
      (26,8.320)  (27,8.740)  (28,9.160) (29,9.300) (30,9.540)
      (31,9.940)  (32,10.280) (33,10.500)(34,10.920)(35,11.020)
      (36,11.140) (37,12.160) (38,12.420)(39,12.660)(40,13.100)
      (41,13.300) (42,13.080) (43,10.040)(44,12.840)(45,14.520)
      (46,14.880) (47,15.020) (48,15.020)(49,17.480)(50,17.280)
      (51,17.740) (52,18.680) (53,19.180)(54,19.700)(55,20.240)
      (56,21.280) (57,22.080) (58,21.200)(59,22.200)(60,22.400)
      (61,23.200) (62,23.600) (63,23.100)(64,24.600)(65,25.400)
      (66,26.000) (67,25.400) (68,26.600)(69,27.200)(70,27.400)
      (71,27.800) (72,28.200) (73,27.400)(74,28.600)(75,29.200)
      (76,30.000) (77,29.400) (78,31.800)(79,32.200)(80,33.800)
      (81,33.400) (82,34.800) (83,34.200)(84,36.400)(85,36.600)
      (86,37.800) (87,36.800) (88,37.600)(89,40.400)(90,41.600)
      (91,40.600) (92,41.800) (93,42.600)(94,42.600)(95,42.400)
      (96,42.200) (97,45.800) (98,45.200)(99,44.600)(100,44.200)
      (101,44.600)(102,44.600)(103,45.800)(104,40.800)(105,39.200)
      (106,39.200)(107,38.800)(108,39.400)(109,41.200)(110,39.200)
      (111,42.000)(112,41.800)(113,42.800)(114,42.000)(115,39.000)
      (116,40.400)(117,41.000)(118,42.800)(119,43.600)(120,42.600)
      (121,44.400)(122,44.900)(123,44.100)(124,43.800)(125,45.800)
      (126,44.600)(127,45.000)(128,45.600)(129,45.800)(130,45.600)
      (131,45.200)(132,44.200)(133,45.800)(134,44.600)(135,45.800)
      (136,46.000)(137,47.600)(138,47.200)(139,49.200)(140,49.000)
      (141,49.800)(142,48.800)(143,49.800)(144,49.400)(145,48.400)
      (146,49.400)(147,50.600)(148,51.800)(149,52.800)(150,44.600)
      (151,45.800)(152,45.600)(153,45.400)(154,45.800)(155,46.000)
      (156,45.200)(157,45.800)(158,45.600)(159,44.400)(160,44.600)
    };
  \end{axis}
\end{tikzpicture}
\end{flushleft}
\pagebreak

Cross-continental connections in North America behave similarly (\texttt{us-west1-b} to \texttt{us-east1-d}):
\begin{flushleft}
\centering  
\begin{tikzpicture}
  \centering
  \begin{axis}[
    xlabel={Number of Concurrent Connections},
    ylabel={Throughput (Gbit/s, TX+RX)},
    grid=major,
    width=12cm,
    height=6cm,
    ymin=0,
  ]
    \addplot[
      thick,
    ] coordinates {
      (1,0.57)   (2,1.178)  (3,1.608)  (4,3.100)  (5,3.780)  (6,4.960)  (7,5.200)  (8,6.440)  (9,6.560)  (10,8.080)
      (11,8.740) (12,10.280) (13,9.340) (14,11.060) (15,11.540) (16,13.260) (17,13.560) (18,14.480) (19,14.720) (20,15.340)
      (21,16.120) (22,17.720) (23,19.360) (24,20.800) (25,20.400) (26,22.400) (27,21.800) (28,23.600) (29,22.200) (30,24.600)
      (31,25.600) (32,25.200) (33,26.200) (34,28.600) (35,28.400) (36,30.400) (37,30.200) (38,30.200) (39,33.000) (40,33.000)
      (41,34.000) (42,34.400) (43,35.800) (44,37.600) (45,36.800) (46,36.600) (47,38.400) (48,39.400) (49,40.400) (50,40.600)
      (51,42.400) (52,42.400) (53,42.200) (54,44.600) (55,44.600) (56,44.800) (57,45.800) (58,46.200) (59,47.800) (60,49.000)
      (61,50.000) (62,50.800) (63,51.200) (64,52.400) (65,51.200) (66,51.400) (67,51.000) (68,52.000) (69,54.400) (70,55.200)
      (71,54.800) (72,54.200) (73,54.600) (74,54.600) (75,52.600) (76,50.600) (77,53.200) (78,55.200) (79,54.200) (80,53.800)
      (81,53.200) (82,54.000) (83,50.800) (84,51.000) (85,54.400) (86,52.200) (87,49.400) (88,49.400) (89,53.800) (90,49.800)
      (91,53.400) (92,54.000) (93,53.600) (94,53.000) (95,51.600) (96,50.400) (97,53.000) (98,53.800) (99,53.600) (100,53.600)
      (101,53.800) (102,53.800) (103,53.400) (104,53.400)
    };
  \end{axis}
\end{tikzpicture}
\end{flushleft}

\subsection{WAN All Reduce Benchmarks}

\subsubsection{Experiment 1: All Reduce Across North America and Europe}
In this experiment, we all reduce 1.073 GB of data across 18 nodes in North America and Europe.
The nodes are connected via Google Cloud's \texttt{Tier\_1} networking performance tier which uses close to optimal BGP routes between regions.
$1024\cdot1024\cdot256=268,435,456$ 32-bit float values are reduced without quantization or compression.
The all reduce takes 90.5 seconds to complete, resulting in an implementation agnostic reduce throughput of 11.85 MB/s computed as the size of the reduce contribution of one peer divided by the time it takes to reduce across all ranks.
The bandwidth utilization per peer (TX + RX bytes transmitted per second) is 358.4 Mbit/s.
It should be noted that this experiment is fatally bottlenecked by the hop over the respective undersea cable connecting the US and Europe.

\begin{figure}[H]
  \centering
  \includegraphics[width=13cm]{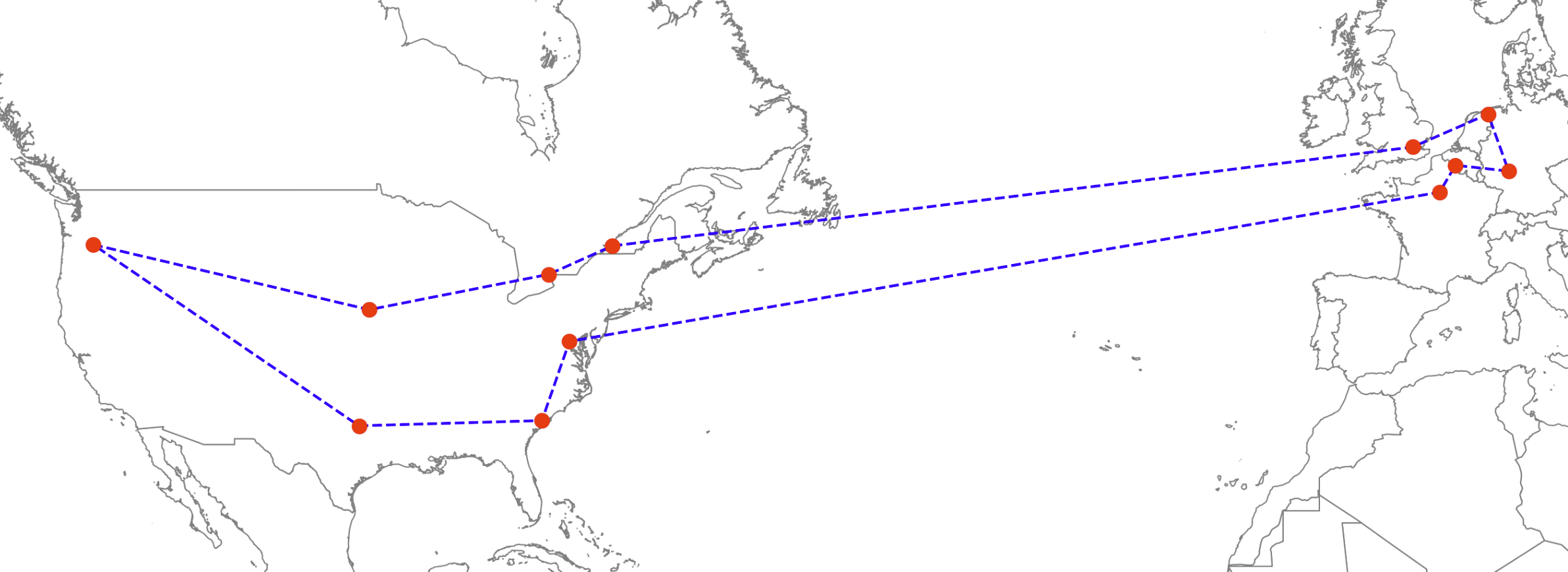}
  \caption{All Reduce Across North America and Europe}
\end{figure}

\begin{table}[H]
  \centering
  \caption{All Reduce Peers and their respective locations}
  \begin{tabular}{llp{5cm}}
    \toprule
    \textbf{Peer} & \textbf{Region} & \textbf{Location} \\
    \midrule
    \multicolumn{3}{c}{\textit{North America}} \\
    \midrule
    peer-01 & us-west1-b & The Dalles, Oregon \\
    peer-02 & us-south1-c & Dallas, Texas \\
    peer-03 & us-east1-d & Moncks Corner, South Carolina \\
    peer-04 & us-central1-c & Council Bluffs, Iowa \\
    peer-10 & northamerica-northeast1-a & Montréal, Québec \\
    peer-11 & northamerica-northeast2-b & Toronto, Ontario \\
    peer-12 & us-central1-c & Council Bluffs, Iowa \\
    peer-14 & us-west1-b & The Dalles, Oregon \\
    peer-15 & us-south1-c & Dallas, Texas \\
    peer-16 & us-east4-a & Ashburn, Virginia \\
    peer-17 & us-east1-d & Moncks Corner, South Carolina \\
    peer-18 & us-east4-a & Ashburn, Virginia \\
    \midrule
    \multicolumn{3}{c}{\textit{Europe}} \\
    \midrule
    peer-05 & europe-west3-c & Frankfurt, Germany \\
    peer-06 & europe-west9-b & Paris, France \\
    peer-07 & europe-west1-b & St. Ghislain, Belgium \\
    peer-08 & europe-west2-b & London, United Kingdom \\
    peer-09 & europe-west4-c & Eemshaven, Netherlands \\
    peer-13 & europe-west1-b & St. Ghislain, Belgium \\
    \bottomrule
  \end{tabular}
\end{table}

\begin{table}[H]
  \centering
  \begin{tabular}{llp{5cm}}
    \toprule
    \textbf{Metric} & \textbf{Value} \\
    \midrule
    World Size & 18 \\
    Reduce Contribution & 1.073 GB \\
    Reduce Contribution Elements & 268,435,456 float32 values \\
    Reduce Time & 90.5 s $\pm$ 0.345s\\
    Effective Reduce Throughput & 11.85 MB/s \\
    Bandwidth Utilization & 358.4 Mbit/s \\
    Tx-Bytes per peer & 2.02818 GB \\
    Rx-Bytes per peer & 2.02818 GB \\
    \bottomrule
  \end{tabular}
  \vspace{0.5cm}
  \caption{Performance Metrics}
\end{table}

\subsubsection{Experiment 1.1: Concurrent Connections North America+Europe}
When using multiple concurrent all reduce operations and dispatching to multiple connections, we observe the following scaling in the setting of the All Reduce North-America \& Europe West experiment:

\begin{table}[H]
  \centering
  \setlength{\tabcolsep}{5pt}
  \caption{Summary: All-Reduce Performance (Europe West, 6 nodes)}  
  \begin{tabularx}{\textwidth}{@{} 
      r  
      *{1}{X}  
      *{1}{X}  
      *{1}{X}  
      *{1}{X}  
    @{}}
    \toprule
    \# CON & Time (s ± s) & eff. TPT (GB/s) & TX+RX\newline BW (Gbit/s) & TX+RX/peer (GB) \\
    \midrule
    \textbf{128} & 10.726 $\pm$ 1.315 & 0.800 & 24.54 $\pm$ 2.923 & 32.45 \\
    \textbf{100} & 10.046 $\pm$ 3.563 & 0.667 & 21.70 $\pm$ 4.774 & 25.35 \\
    \textbf{64} & 11.338 $\pm$ 1.870 & 0.379 & 11.73 $\pm$ 1.854 & 16.22 \\
    \textbf{32} & 7.614 $\pm$ 0.608 & 0.282 & 8.57 $\pm$ 0.654 & 8.11 \\
    \textbf{16} & 9.900 $\pm$ 2.511 & 0.108 & 3.43 $\pm$ 0.664 & 4.05 \\
    \textbf{8} & 4.356 $\pm$ 0.474 & 0.123 & 1.88 $\pm$ 0.207 & 1.01 \\
    \bottomrule
  \end{tabularx}
\end{table}

\begin{flushleft}
  \centering
  \begin{tikzpicture}
    \begin{axis}[
        xlabel={Number of Concurrent Connections},
        ylabel={Bandwidth Utilization (Gbit/s)},
        grid=major,
    ]
    \addplot[
      mark=o,
      thick,
      error bars/.cd,
        y dir=both,
        y explicit,
    ] coordinates {
      (8, 1.88) +- (0, 0.207)
      (16, 3.43) +- (0, 0.664)
      (32, 8.57) +- (0, 0.654)
      (64, 11.73) +- (0, 1.854)
      (100, 21.70) +- (0, 4.774)
      (128, 24.54) +- (0, 2.923)
    };
    \end{axis}
  \end{tikzpicture}
\end{flushleft}

\begin{table}[H]
  \centering
  \begin{tabular}{ll}
    \toprule
    \textbf{Metric} & \textbf{Value} \\
    \midrule
    Number of Peers & 18 \\
    Reduce Contribution per Reduce Operation & 64 MiB \\
    Number of Reduce Operations & \{ 8, 16, 32, 64, 100, 128 \} \\
    Max Number of Concurrent Reduce Operations & \{ 8, 16, 32, 64, 100, 128 \} \\
    P2P Connection Pool Size & \{ 8, 16, 32, 64, 100, 128 \} \\
    \bottomrule
  \end{tabular}
  \vspace{0.5cm}
  \caption{Concurrent All-Reduce Experiment Details}
\end{table}

\subsection{Experiment 2: All Reduce Across North America}
Experiment 2 is a reduced version of Experiment 1, where we all reduce 1.073 GB of data across 12 nodes in North America.

\begin{figure}[H]
  \centering
  \includegraphics[width=13cm]{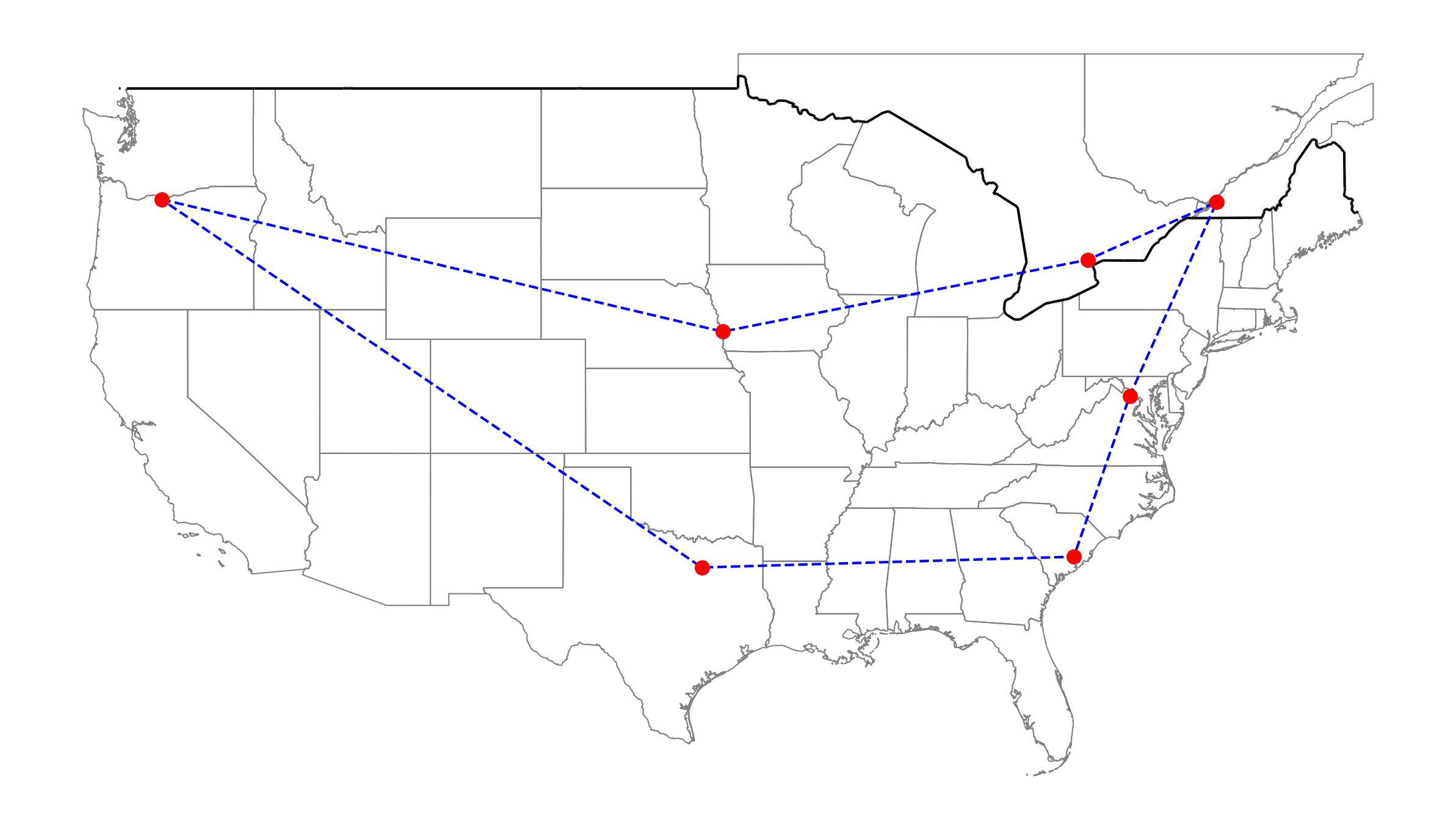}
  \caption{All Reduce Across North America}
\end{figure}

\begin{table}[H]
  \centering
  \caption{All Reduce Peers and their respective locations}
  \begin{tabular}{llp{5cm}}
    \toprule
    \textbf{Peer} & \textbf{Region} & \textbf{Location} \\
    \midrule
    \multicolumn{3}{c}{\textit{North America}} \\
    \midrule
    peer-01 & us-west1-b & The Dalles, Oregon \\
    peer-02 & us-south1-c & Dallas, Texas \\
    peer-03 & us-east1-d & Moncks Corner, South Carolina \\
    peer-04 & us-central1-c & Council Bluffs, Iowa \\
    peer-05 & northamerica-northeast1-a & Montréal, Québec \\
    peer-06 & northamerica-northeast2-b & Toronto, Ontario \\
    peer-07 & us-central1-c & Council Bluffs, Iowa \\
    peer-08 & us-west1-b & The Dalles, Oregon \\
    peer-09 & us-south1-c & Dallas, Texas \\
    peer-10 & us-east4-a & Ashburn, Virginia \\
    peer-11 & us-east1-d & Moncks Corner, South Carolina \\
    peer-12 & us-east4-a & Ashburn, Virginia \\
    \bottomrule
  \end{tabular}
\end{table}
\begin{table}[H]
  \centering
  \begin{tabular}{llp{5cm}}
    \toprule
    \textbf{Metric} & \textbf{Value} \\
    \midrule
    World Size & 12 \\
    Reduce Contribution & 1.073 GB \\
    Reduce Contribution Elements & 268,435,456 float32 values \\
    Reduce Time & 35.2 s $\pm$ 0.306s \\
    Effective Reduce Throughput & 30.48 MB/s \\
    Bandwidth Utilization & 897.616 Mbit/s \\
    Tx-Bytes per peer & 1.9685 GB \\
    Rx-Bytes per peer & 1.9685 GB \\
    \bottomrule
  \end{tabular}
  \vspace{0.5cm}
  \caption{Performance Metrics}
\end{table}

\subsubsection{Experiment 2.1: Concurrent Connections North America}
When using multiple concurrent all reduce operations and dispatching to multiple connections, we observe the following scaling in the setting of the All Reduce North-America experiment.

\begin{table}[H]
  \centering
  \setlength{\tabcolsep}{5pt}
  \caption{Summary: All-Reduce Performance (North America, 12 nodes)}  
  \begin{tabularx}{\textwidth}{@{} 
      r  
      *{1}{X}  
      *{1}{X}  
      *{1}{X}  
      *{1}{X}  
    @{}}
    \toprule
    \# CON & Time (s ± s) & eff. TPT (GB/s) & TX+RX\newline BW (Gbit/s) & TX+RX/peer (GB) \\
    \midrule
    \textbf{128} & 9.419 $\pm$ 1.753 & 0.911 & 27.57 $\pm$ 4.822 & 31.49 \\
    \textbf{64} & 4.894 $\pm$ 0.597 & 0.878 & 26.08 $\pm$ 2.948 & 15.74 \\
    \textbf{32} & 3.350 $\pm$ 0.179 & 0.641 & 18.85 $\pm$ 0.997 & 7.87 \\
    \textbf{16} & 3.572 $\pm$ 0.705 & 0.299 & 9.01 $\pm$ 0.125 & 3.93 \\
    \textbf{8} & 3.227 $\pm$ 0.564 & 0.166 & 4.88 $\pm$ 0.010 & 1.96 \\
    \bottomrule
  \end{tabularx}
\end{table}

\begin{flushleft}
  \centering
  \begin{tikzpicture}
    \begin{axis}[
        xlabel={Number of Concurrent Connections},
        ylabel={Bandwidth Utilization (Gbit/s)},
        grid=major,
    ]
    \addplot[
      mark=o,
      thick,
      error bars/.cd,
        y dir=both,
        y explicit,
    ] coordinates {
      (8, 4.88) +- (0, 0.010)
      (16, 9.01) +- (0, 0.125)
      (32, 18.85) +- (0, 0.997)
      (64, 26.08) +- (0, 2.948)
      (128, 27.57) +- (0, 4.822)
    };
    \end{axis}
  \end{tikzpicture}
\end{flushleft}

\begin{table}[H]
  \centering
  \begin{tabular}{ll}
    \toprule
    \textbf{Metric} & \textbf{Value} \\
    \midrule
    Number of Peers & 12 \\
    Reduce Contribution per Reduce Operation & 64 MiB \\
    Number of Reduce Operations & \{ 8, 16, 32, 64, 128 \} \\
    Max Number of Concurrent Reduce Operations & \{ 8, 16, 32, 64, 128 \} \\
    P2P Connection Pool Size & \{ 8, 16, 32, 64, 128 \} \\
    \bottomrule
  \end{tabular}
  \vspace{0.5cm}
  \caption{Concurrent All-Reduce Experiment Details}
\end{table}

\subsection{Experiment 3: All Reduce Europe West}
In this experiment we all reduce 1.073 GB of data across 6 nodes in western Europe - analogous to Experiment 3 of the previous section - while scaling the number of concurrent all reduce operations.

\begin{figure}[H]
  \centering
  \includegraphics[width=10cm]{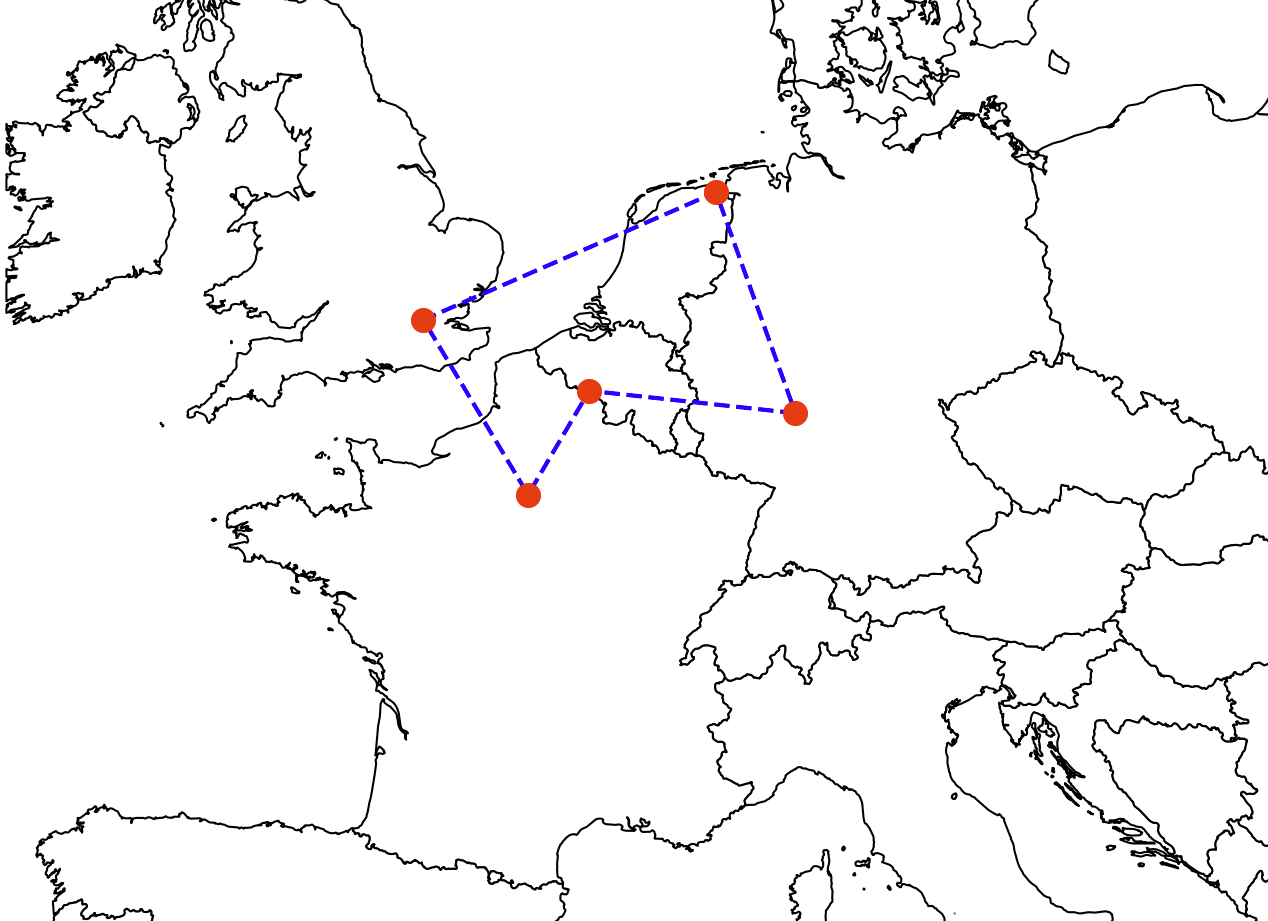}
  \caption{All Reduce Europe West}
\end{figure}

\begin{table}[htbp]
  \centering
  \caption{All Reduce Peers and their respective locations}
  \begin{tabular}{llp{5cm}}
    \toprule
    \textbf{Peer} & \textbf{Region} & \textbf{Location} \\
    \midrule
    \multicolumn{3}{c}{\textit{Europe}} \\
    \midrule
    peer-01 & europe-west3-c & Frankfurt, Germany \\
    peer-02 & europe-west9-b & Paris, France \\
    peer-03 & europe-west1-b & St. Ghislain, Belgium \\
    peer-04 & europe-west2-b & London, United Kingdom \\
    peer-05 & europe-west4-c & Eemshaven, Netherlands \\
    peer-06 & europe-west1-b & St. Ghislain, Belgium \\
    \bottomrule
  \end{tabular}
\end{table}
\begin{table}[H]
  \centering
  \begin{tabular}{llp{5cm}}
    \toprule
    \textbf{Metric} & \textbf{Value} \\
    \midrule
    World Size & 6 \\
    Reduce Contribution & 1.073 GB \\
    Reduce Contribution Elements & 268,435,456 float32 values \\
    Reduce Time & 8.3 s $\pm$ 0.328s \\
    Effective Reduce Throughput & 129.2 MB/s \\
    Bandwidth Utilization & 3.6704 Gbit/s \\
    Tx-Bytes per peer & 1.7895 GB \\
    Rx-Bytes per peer & 1.7895 GB \\
    \bottomrule
  \end{tabular}
\end{table}

\subsubsection{Experiment 3.1: Concurrent Connections Europe West}
When using multiple concurrent all reduce operations and dispatching to multiple connections, we observe the following scaling in the setting of the Europe-West experiment:

\begin{table}[H]
  \centering
  \setlength{\tabcolsep}{5pt}
  \caption{Summary: All-Reduce Performance (Europe West, 6 nodes)}  
  \begin{tabularx}{\textwidth}{@{} 
      r  
      *{1}{X}  
      *{1}{X}  
      *{1}{X}  
      *{1}{X}  
    @{}}
    \toprule
    \# CON & Time (s ± s) & eff. TPT (GB/s) & TX+RX\newline BW (Gbit/s) & TX+RX/peer (GB) \\
    \midrule
    \textbf{128} & 5.066 $\pm$ 0.560 & 1.694 & 45.74 $\pm$ 5.265 & 28.63 \\
    \textbf{100} & 4.136 $\pm$ 0.757 & 1.622 & 44.55 $\pm$ 7.733 & 22.36 \\
    \textbf{64} & 2.596 $\pm$ 0.233 & 1.655 & 44.47 $\pm$ 3.964 & 14.31 \\
    \textbf{32} & 1.684 $\pm$ 0.167 & 1.275 & 34.29 $\pm$ 3.127 & 9.21 \\
    \textbf{16} & 1.602 $\pm$ 0.219 & 0.669 & 18.17 $\pm$ 2.321 & 5.04 \\
    \textbf{8} & 1.292 $\pm$ 0.159 & 0.415 & 11.25 $\pm$ 1.398 & 3.11 \\
    \bottomrule
  \end{tabularx}
\end{table}

\begin{flushleft}
  \centering
  \begin{tikzpicture}
    \begin{axis}[
        xlabel={Number of Concurrent Connections},
        ylabel={Bandwidth Utilization (Gbit/s)},
        grid=major,
    ]
    \addplot[
      mark=o,
      thick,
      error bars/.cd,
        y dir=both,
        y explicit,
    ] coordinates {
      (8, 11.25) +- (0, 1.398)
      (16, 18.17) +- (0, 2.321)
      (32, 34.29) +- (0, 3.127)
      (64, 44.47) +- (0, 3.964)
      (100, 44.55) +- (0, 7.733)
      (128, 45.74) +- (0, 5.265)
    };
    \end{axis}
  \end{tikzpicture}
\end{flushleft}

\begin{table}[H]
  \centering
  \begin{tabular}{ll}
    \toprule
    \textbf{Metric} & \textbf{Value} \\
    \midrule
    Number of Peers & 12 \\
    Reduce Contribution per Reduce Operation & 64 MiB \\
    Number of Reduce Operations & \{ 8, 16, 32, 64, 128 \} \\
    Max Number of Concurrent Reduce Operations & \{ 8, 16, 32, 64, 128 \} \\
    P2P Connection Pool Size & \{ 8, 16, 32, 64, 128 \} \\
    \bottomrule
  \end{tabular}
  \vspace{0.5cm}
  \caption{Concurrent All-Reduce Experiment Details}
\end{table}

\subsection{Comparisons with Gloo (without VPN over WAN)}
It should be noted that Gloo is not designed to operate over the public internet, however given that all of our nodes have internal IP addresses assigned by our cloud provider, we can measure
Gloo's performance over the public internet without the overhead of a client-side software VPN.
For this comparison, we replicate the Experiments 1, 2 and 3 with Gloo on the same nodes.
Across the board, we observe a gap in performance of Gloo compared to PCCL, which we attribute to lack of topology optimization, resulting in a naive rank launching order propagating into a suboptimal ring order.

\begin{table}[H]
  \centering
  \caption{Comparison of reduce time between PCCL and Gloo}
  \begin{tabular}{lccc}
    \toprule
    \textbf{Experiment} & \textbf{PCCL Time (s)} & \textbf{Gloo Time (s)} & \textbf{Improvement (\%)} \\
    \midrule
    \textbf{1.} North America + Europe & 90.50 $\pm$ 0.35 & 94.44 $\pm$ 1.84 & 4.15\% \\
    \textbf{2.} North America & 35.20 $\pm$ 0.31 & 37.58 $\pm$ 0.85 & 6.33\% \\
    \textbf{3.} Europe & 8.30  $\pm$ 0.33 & 9.67  $\pm$ 0.77 & 14.17\% \\
    \bottomrule
  \end{tabular}
\end{table}

\begin{table}[H]
  \centering
  \caption{Comparison of effective throughput between PCCL and Gloo}
  \begin{tabular}{lccc}
    \toprule
    \textbf{Experiment} & \textbf{PCCL eff. TPT (MB/s)} & \textbf{Gloo eff. TPT (MB/s)} \\
    \midrule
    \textbf{1.} North America + Europe & 11.85 & 11.37 \\
    \textbf{2.} North America & 30.48 & 28.82 \\
    \textbf{3.} Europe & 129.20 & 113.66 \\
    \bottomrule
  \end{tabular}
\end{table}

It should be noted that Gloo does not natively support concurrent all-reduce operations and can thus not effectively utilize the most crucial trick to maximize throughput over the public internet.

\section{Related Work}
\paragraph{Gloo \citep{Gloo}}  
Gloo is a collective library optimized for on-premise co-located clusters. Because it assumes full peer reachability over a secure local network, it cannot natively be used over the public internet without employing a VPN. This significantly holds back Gloo in this regard and is a major reason why PCCL was developed.
This was our approach for INTELLECT-1 \citep{intellect1}, which resulted infrastructure that required heavy manual intervention/on-boarding steps instead of an automatically handled, robust join procedure.

\paragraph{Hivemind \citep{hivemind}} 
Hivemind is a decentralized collective communication and training framework built on top of PyTorch that uses a gossip-based peer sampling service and a distributed hash table (DHT) for peer discovery and parameter exchange. Unlike PCCL's master--client micro-consensus model, Hivemind operates in a fully decentralized fashion: peers form dynamic shards via Rendezvous hashing, performing asynchronous, compressed all-reduce rounds within each shard. Hivemind employs gradient compressors (e.g., Top-K sparsification, signSGD) with error-feedback loops to reduce bandwidth, and relies on best-effort RPC (e.g., gRPC) and timeouts to handle peer churn. This design allows flexible on/off‐ramping of workers without global coordination, but sacrifices strict synchronization: collective results converge only eventually rather than bit‐exactly, making deterministic state progression across heterogeneous hardware unattainable.

\paragraph{Horovod + Elastic} 
Elastic Horovod extends Horovod's \citep{sergeev2018horovodfasteasydistributed} MPI/Gloo collectives with a Python-level rendezvous and in-memory rollback API that lets workers join or depart without a full job restart. Horovod+Elastic drives recovery from failed collective operations through Python exceptions via an explicit rollback.  While this makes moderate-scale elasticity easy to script, it also:

\begin{itemize}
  \item Incurs non-trivial latency from rollbacks and recovery (potentially seconds)
  \item Relies on application-level checkpoint hooks rather than a low-level micro-consensus, leaving subtle failure paths largely unexercised,
  \item Does not guarantee bit-identical state across ranks — merely logical equivalence via user-supplied state objects.
\end{itemize}

In contrast, PCCL is fault tolerant in and of itself as opposed to wrapping an inherently non-fault tolerant lower-level library, uses a fine-grained consensus protocol, facilitating a design where possible error paths are enumerable and testable, and can enforce bit-exact shared-state progression purely from collective-op results. PCCL's error paths in comparison are roughly equally fast as the success paths and do not require dedicated rollback mechanisms.

\section{Practical Implications}
With PCCL's strong fault tolerance and dynamic membership, PCCL makes training on spot instances practically feasible.
With only one non-spot instance, the run is guaranteed stable and loss of training state is impossible.
Additionally, PCCL will allow for aggregation of compute resources across the internet with previously infeasible ease, resulting in less idle compute in underutilized data centers.
Thus, Hybrid- and Multi-Cloud Training or even Edge and Federated Learning for less communication-intensive training workloads will be easy to orchestrate.
In addition, PCCL's shared state system implicitly protects model weights and optimizer state, where even if all peers have dropped from the training run, resumption will only be allowed from the previous shared state hash, which peers are expected to restore from local checkpoints. This categorically rules out training-dynamics inconsistencies introduced by restarts assuming an i.i.d data loader.

\section{Conclusion}

In this work, we presented the Prime Collective Communications Library (\pccl), a novel, fault-tolerant collective communications library tailored for dynamic, wide-area distributed machine learning workloads. Unlike traditional MPI-derived libraries such as NCCL, \pccl{} natively supports peer churn through a master-client model and fine-grained micro-consensus, enabling peers to join or fail without restarting the entire training job. We detailed how \pccl{} maintains bit-identical shared state via parallel hashing and on-demand data transfer, optimizes ring topologies through p2p bandwidth measurements and ATSP solvers, and provides interruptible, zero-copy pipelined collectives. Through extensive stress tests across Linux, macOS, and Windows—subjecting the system to rapid peer churn and concurrent operations—and implementation-agnostic throughput benchmarks, we demonstrated that \pccl{} is both robust and competitive with existing libraries.
Beyond the core library, we showed how \pccl{} enables the straightforward implementation of communication-efficient training algorithms such as DiLoCo and its asynchronous variant, further reducing bandwidth requirements and hiding communication latency behind compute. These examples illustrate \pccl's versatility as a low-level primitive for both classical DDP and next-generation low-communication optimizers over unreliable or heterogeneous networks.
We also aim to investigate adaptive synchronization schedules and incremental topology optimization driven by real-time telemetry. By lowering the barrier to resilient, internet-scale training, \pccl{} opens the door to new distributed ML workflows on spot instances and across geographically dispersed resources.

\bibliographystyle{plainnat}
\bibliography{references}
\pagebreak

\appendix
\section{Appendix}

\subsection{SimpleHash vs Thrust}
\label{sec:appendix_simplehash_vs_thrust}
In this section we compare the performance of SimpleHash memory-throughput to that of Thrust's reduce kernels.

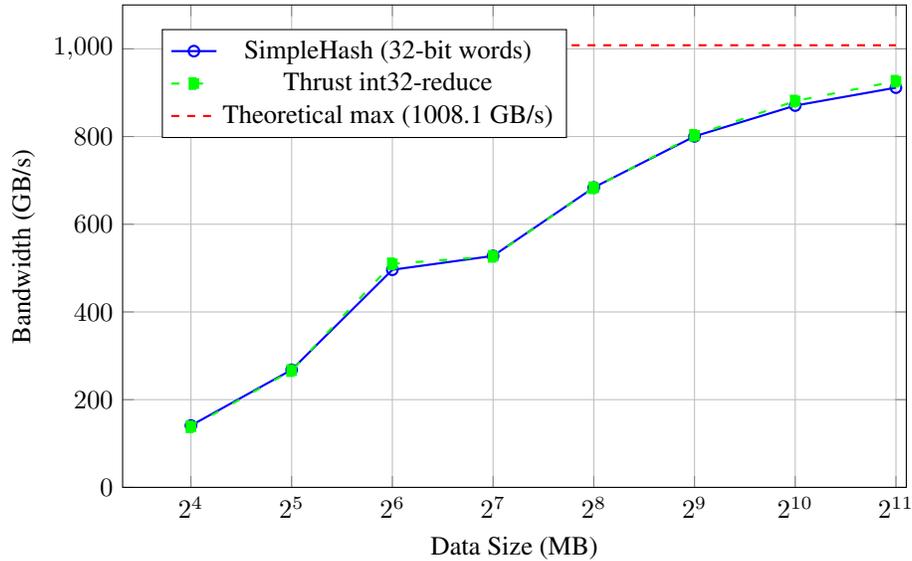
\begin{figure}[H]
    \centering
    \begin{tikzpicture}
      \begin{axis}[
        xlabel={Data Size (MB)},
        ylabel={Bandwidth (GB/s)},
        width=12cm,
        height=8cm,
        grid=both,
        xmin=10,   xmax=2200,
        ymin=0,    ymax=1100,
        xmode=log,
        log basis x=2,
        xtick={16,32,64,128,256,512,1024,2048},
        xticklabel style={/pgf/number format/fixed},
        ytick={0,200,400,600,800,1000},
        legend style={at={(0.05,0.95)}, anchor=north west},
      ]
        \addplot[
          color=blue,
          mark=o,
          thick,
        ] coordinates {
          (16, 140.953)
          (32, 268.027)
          (64, 496.167)
          (128, 527.806)
          (256, 683.747)
          (512, 800.395)
          (1024, 870.564)
          (2048, 911.428)
        };
        \addlegendentry{SimpleHash (32-bit words)}
    
        \addplot[
          color=green,
          mark=square*,
          thick,
          loosely dashed,
        ] coordinates {
          (16, 138.202)
          (32, 266.235)
          (64, 509.804)
          (128, 526.731)
          (256, 683.147)
          (512, 802.716)
          (1024, 880.945)
          (2048, 926.084)
        };
        \addlegendentry{Thrust int32-reduce}
    
        \addplot[
          color=red,
          dashed,
          thick,
        ] coordinates {
          (16,1008.1)
          (2048,1008.1)
        };
        \addlegendentry{Theoretical max (1008.1 GB/s)}
      \end{axis}
    \end{tikzpicture}
    \caption{RTX 4090: SimpleHash vs. Thrust int32 Reduction}
\end{figure}

\subsection{Comparing behavior of \texttt{\_\_nv\_expf} across different architectures}
\label{sec:appendix_expf_determinism}

We compare the behavior of \texttt{\_\_nv\_expf} across different architectures using the following test which hashes the output bits of \texttt{\_\_nv\_expf} across all 2$^{32}$ bit-patterns of float.
\begin{lstlisting}[language=C]
#include <cstdio>
#include <cstdint>
#include <cuda_runtime.h>
#include <cmath>

__global__ void testExpAndHash(unsigned long long* d_nan,
                               unsigned long long* d_inf,
                               unsigned long long* d_finite,
                               unsigned long long* d_sumBits,
                               unsigned int*        d_xorBits)
{
    const unsigned long long N = 0x1ULL << 32;
    unsigned long long stride = gridDim.x * blockDim.x;
    unsigned long long idx    = blockIdx.x * blockDim.x + threadIdx.x;
    unsigned long long nanCt = 0, infCt = 0, finCt = 0;
    unsigned long long sumBits = 0;
    unsigned int    xorBits = 0;
    for (unsigned long long i = idx; i < N; i += stride) {
        float x = __uint_as_float((unsigned int)i);
        float y = expf(x);
        if (isnan(y)) {
            nanCt++;
        } else if (isinf(y)) {
            infCt++;
        } else {
            finCt++;
        }
        unsigned int ybits = __float_as_uint(y);
        sumBits += (unsigned long long)ybits;
        xorBits ^= ybits;
    }
    atomicAdd(d_nan,    nanCt);
    atomicAdd(d_inf,    infCt);
    atomicAdd(d_finite, finCt);
    atomicAdd(d_sumBits, sumBits);
    atomicXor(d_xorBits, xorBits);
}

int main() {
    unsigned long long *d_nan, *d_inf, *d_finite, *d_sumBits;
    unsigned int       *d_xorBits;

    cudaMalloc(&d_nan,      sizeof(unsigned long long));
    cudaMalloc(&d_inf,      sizeof(unsigned long long));
    cudaMalloc(&d_finite,   sizeof(unsigned long long));
    cudaMalloc(&d_sumBits,  sizeof(unsigned long long));
    cudaMalloc(&d_xorBits,  sizeof(unsigned int));

    cudaMemset(d_nan,      0, sizeof(unsigned long long));
    cudaMemset(d_inf,      0, sizeof(unsigned long long));
    cudaMemset(d_finite,   0, sizeof(unsigned long long));
    cudaMemset(d_sumBits,  0, sizeof(unsigned long long));
    cudaMemset(d_xorBits,  0, sizeof(unsigned int));

    const int threadsPerBlock = 256;
    const int blocks          = 1024;

    printf("Launching %d blocks of %d threads\n", blocks, threadsPerBlock);
    testExpAndHash<<<blocks, threadsPerBlock>>>(
        d_nan, d_inf, d_finite, d_sumBits, d_xorBits);
    cudaDeviceSynchronize();

    unsigned long long h_nan, h_inf, h_finite, h_sumBits;
    unsigned int       h_xorBits;
    cudaMemcpy(&h_nan,     d_nan,     sizeof(h_nan),     cudaMemcpyDeviceToHost);
    cudaMemcpy(&h_inf,     d_inf,     sizeof(h_inf),     cudaMemcpyDeviceToHost);
    cudaMemcpy(&h_finite,  d_finite,  sizeof(h_finite),  cudaMemcpyDeviceToHost);
    cudaMemcpy(&h_sumBits, d_sumBits, sizeof(h_sumBits), cudaMemcpyDeviceToHost);
    cudaMemcpy(&h_xorBits, d_xorBits, sizeof(h_xorBits), cudaMemcpyDeviceToHost);

    cudaFree(d_nan);
    cudaFree(d_inf);
    cudaFree(d_finite);
    cudaFree(d_sumBits);
    cudaFree(d_xorBits);

    printf("Results over all %llu float bit-patterns:\n", total);
    printf("NaN outputs: %llu\n", h_nan);
    printf("Inf outputs: %llu\n", h_inf);
    printf("Finite outputs: %llu\n", h_finite);
    printf("Sum check: %llu\n", h_nan + h_inf + h_finite);
    printf("Output-bits hash:\n");
    printf("Sum of bits: %llu\n", h_sumBits);
    printf("XOR of bits: 0x%08X\n", h_xorBits);
    return 0;
}
\end{lstlisting}

This results in the following output on all tested architectures:

\begin{verbatim}
Results over all 4294967296 float bit-patterns:
  NaN outputs:    16777214
  Inf outputs:    1020169705
  Finite outputs: 3258020377
  Sum check:      4294967296
Output-bits hash:
  Sum of bits:    4602279786742895247
  XOR of bits:    0x45ABDA35
\end{verbatim}

Architectures tested:
\begin{itemize}
  \item GTX 980 Ti (sm\_52)
  \item GTX 1060 (sm\_61)
  \item RTX 4090 (sm\_89)
  \item GH 200 (sm\_90)
  \item B200 (sm\_100)
\end{itemize}

\end{document}